\documentclass[sigconf]{acmart}
\AtBeginDocument{%
  }

\copyrightyear{2026}
\acmYear{2026}
\setcopyright{cc}
\setcctype{by}
\acmConference[CHI '26]{Proceedings of the 2026 CHI Conference on Human Factors in Computing Systems}{April 13--17, 2026}{Barcelona, Spain}
\acmBooktitle{Proceedings of the 2026 CHI Conference on Human Factors in Computing Systems (CHI '26), April 13--17, 2026, Barcelona, Spain}
\acmPrice{}
\acmDOI{10.1145/3772318.3791787}
\acmISBN{979-8-4007-2278-3/2026/04}

\begin{document}

\title{Towards Fluent Interaction with Cyber-Physical Architecture}


\author{Jesse T. Gonzalez}
\affiliation{%
  \institution{Carnegie Mellon University}
  \city{Pittsburgh}
  \country{USA}}
\email{jtgonzal@cs.cmu.edu}

\author{Neeta Khanuja}
\affiliation{%
  \institution{CMU, Instituto Superior Técnico}
  \city{Lisbon}
  \country{Portugal}}
\email{nkhanuja@andrew.cmu.edu }

\author{Michael Li}
\affiliation{%
  \institution{Carnegie Mellon University}
  \city{Pittsburgh}
  \country{USA}}
\email{mml2@andrew.cmu.edu}

\author{Maggie Guo}
\affiliation{%
  \institution{Carnegie Mellon University}
  \city{Pittsburgh}
  \country{USA}}
\email{yuetongg@andrew.cmu.edu}

\author{Layomi Olaitan}
\affiliation{%
  \institution{Carnegie Mellon University}
  \city{Pittsburgh}
  \country{USA}}
\email{lolaitan@andrew.cmu.edu}

\author{Emily Lau}
\affiliation{%
  \institution{Carnegie Mellon University}
  \city{Pittsburgh}
  \country{USA}}
\email{emilylau@andrew.cmu.edu}

\author{Jenny Pugh}
\affiliation{%
  \institution{Carnegie Mellon University}
  \city{Pittsburgh}
  \country{USA}}
\email{jpugh@andrew.cmu.edu}

\author{Alexandra Ion}
\affiliation{%
  \institution{Carnegie Mellon University}
  \city{Pittsburgh}
  \country{USA}}
\email{alexandraion@cmu.edu}

\author{Scott E. Hudson}
\affiliation{%
  \institution{Carnegie Mellon University}
  \city{Pittsburgh}
  \country{USA}}
\email{scott.hudson@cs.cmu.edu}

\renewcommand{\shortauthors}{Gonzalez et al.}

\begin{abstract}
What happens when your walls begin to move? This paper explores the design of human-robot interaction for architectural-scale, shape-changing environments. We present findings from two studies: (1) a series of speculative design workshops (N=20) that uncovered aspirational visions for these spaces, and (2) a task-based Wizard-of-Oz elicitation study (N=12) that grounded these visions in the challenges of practical interaction. Our workshop findings reveal a complex landscape of user desires, exposing critical tensions between proactive automation and the preservation of user autonomy, and between personalization and public ownership. Our elicitation study reveals a set of core interaction challenges related to multimodal collaboration; and, most critically: suggests the need for a modality-agnostic model of evolving user intent. We conclude with a set of grounded proposals for creating robotic environments that are collaborative and trusted partners in everyday life.
\end{abstract}


\begin{teaserfigure}
  \includegraphics[width=\textwidth]{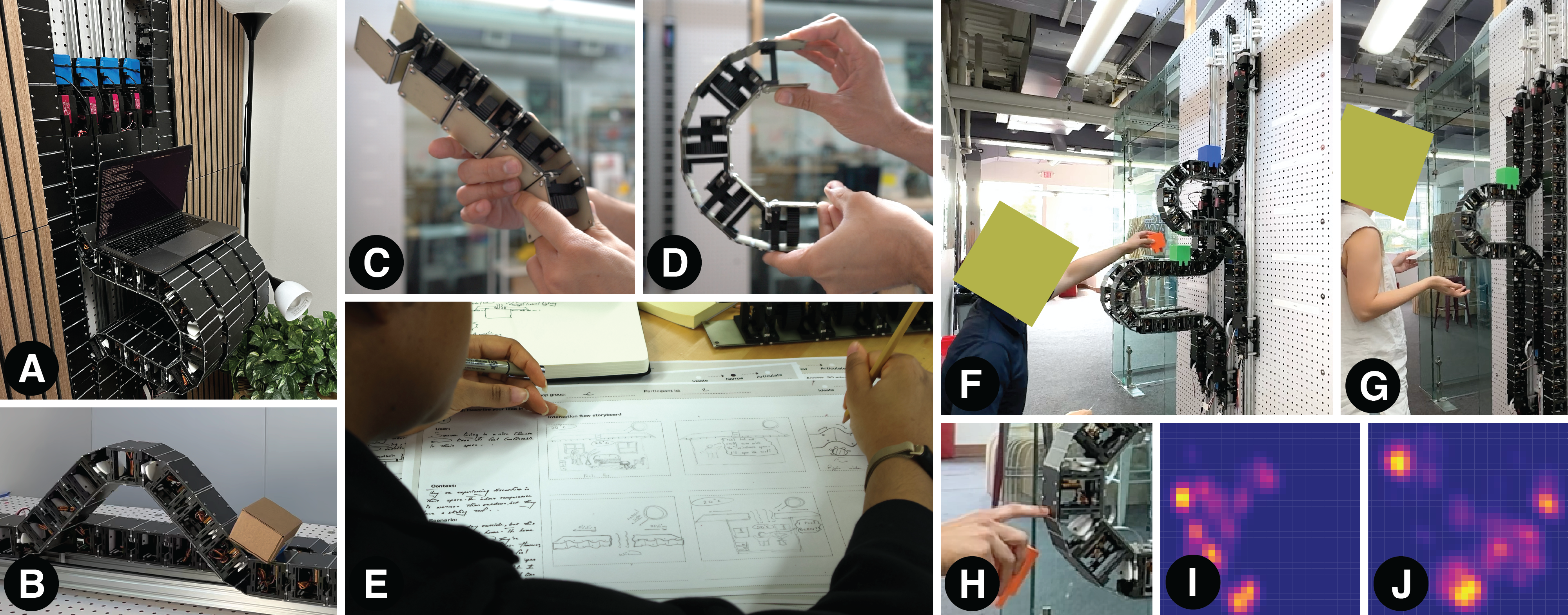}
  \caption{Shape-changing walls and floors (a,b) afford exciting new interactions with the built environment. To uncover the most compelling contexts, and the human values that underlie them, we conducted a series of speculative design workshops (c-e) where architects and designers received hands-on experience observing and ideating upon these technologies.  We followed this with an elicitation study (f,g), analyzing the multimodal interaction strategies (h-j) that arose during collaborative tasks with our robotic wall.}
  \Description{...}
  \label{fig:teaser}
\end{teaserfigure}

\maketitle


\section{Introduction}

Since constructing the first shelters, we have largely demanded \emph{rigidity} from the built environment — we expect it to be static. But if we instead entertained the idea that the walls around us could adapt and transform (without compromising stability), then what would we desire from these shape-changing surroundings? What new capabilities could they afford? And how would we, as ``users'', communicate our intentions to architectural elements that have become, in some sense, autonomous agents?

\begin{figure}[b]
  \centering
  \includegraphics[width=8.45cm]{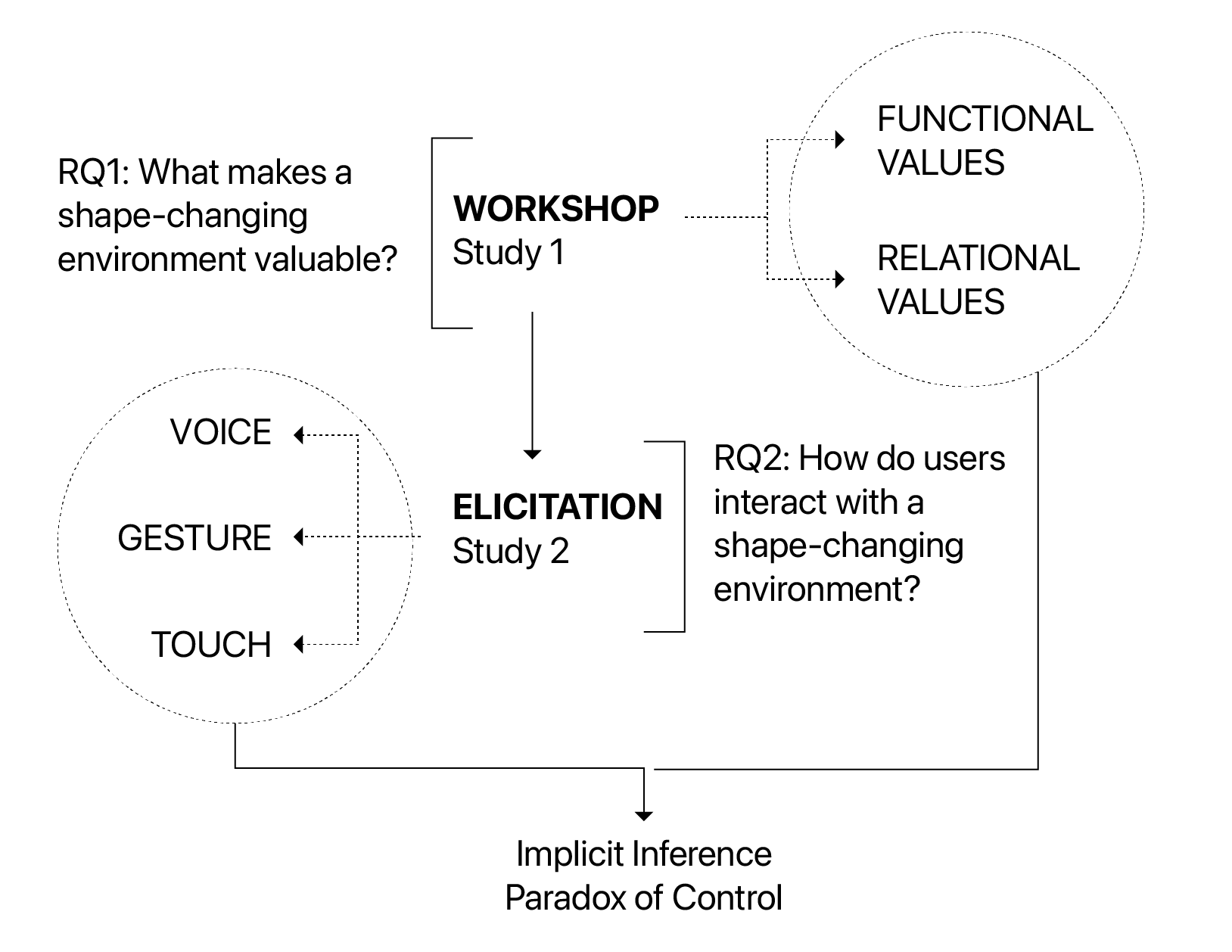}
  \caption{Through two complementary studies, we sought to understand what people might desire from shape-changing architecture, and how they interact with such systems in practice. Both studies point to a tension relating to implicit context: to allow for the types naturalistic interactions that were envisioned by participants in Study 1 and enacted by participants in Study 2, robotic environments may need to provocatively infer instructions that users do not explicitly command.}
  \Description{A diagram shows how two the research questions, RQ1 and RQ2, are connected to the design workshop and elicitation study respectively.}
\end{figure}

As robotic technologies transcend the industrial sector — entering homes \cite{shafiullah2023bringing, forlizzi2007robotic, Metz2025HumanoidRobots}, offices \cite{stock2023one, mutlu2008robots, YoonWakabayashi2022NaverRobots}, and public spaces \cite{weinberg2023sharing, brown2024trash, seyitouglu2022understanding, kamino2023coffee} — these questions become more salient. While one strain of futurists envisions a world that is brimming with roving automatons, a smaller community champions a more “unobtrusive” vision of ubiquitous robotics \cite{Han2025UnobtrusiveAI, wang2021space}; one in which users benefit from automation embedded within the built environment itself. Shape-changing surfaces \cite{houayek2014awe, gonzalez2023constraint}, actuated furniture \cite{vazquez2014spatial, kinch2014encounters}, and kinetic facades \cite{sharaidin2024kinetic} promise spaces that can physically adapt to the demands of an increasingly urbanized world. A room that hosts focused work in the morning, for instance, may need to accommodate exercise, dining, or childcare by evening \cite{huynh2022tidal}. A staircase that posed no issue at age forty, may later become an obstacle at age eighty. In such cases, adaptive architecture could assist — transforming when needed, and disappearing into the periphery when not in use.

Unlike smartphones or laptops that we explicitly engage with and then put away, architectural-scale robotic systems are in effect omnipresent — we exist \emph{within} them rather than using them in isolation \cite{kumar2024design}. They must navigate a delicate balance: remaining inconspicuous when at rest, yet being immediately responsive when called upon. They must interpret ambiguous human signals (the placement of an object, a muttered complaint about clutter) and determine whether and how to act. Most importantly, they must do this while preserving human autonomy, building trust through repeated interactions in both public and private spaces.

Such considerations lead us to two questions:

\begin{itemize}
    \item \textbf{RQ1:} What makes a shape-changing environment valuable?
    \item \textbf{RQ2:} How do users naturally interact with a shape-changing environment?
\end{itemize}

This paper addresses these challenges through two complementary studies, examining both the aspirational values and practical realities of interacting with cyber-physical architecture (Figure 2). First, we conduct a series of speculative design workshops to uncover what merit users might find in a shape-changing environment (Figure 1e). Then, we ground these visions in a task-based elicitation study, revealing how users naturally interact with — and recover from errors made by — these systems (Figure 1f).

Our platform for this exploration is a shape-changing robotic wall (Figure 1a). This structure, which we have adapted from \cite{gonzalez2023constraint}, consists of three independent columns, which can move vertically on a set of two-meter rails. By remotely configuring the mechanical constraints that exist within each column, the structure can transition from a flat, rigid state into a curved, protruding shape (i.e. a shelf or an awning). These resulting shapes are stable, and when used as shelves, can hold common household objects such as laptops, backpacks, cookware, and books.

Prior work \cite{gonzalez2023constraint} introduced methods for fabricating and actuating such surfaces. However, user interaction techniques — and design considerations for cyber-physical architecture more broadly — were beyond that initial scope. Our work augments that research, providing a framework for understanding how users conceptualize and communicate with robotic environments.

Specifically, we contribute:

\begin{enumerate}

    \item Eleven design values for shape-changing environments, derived from speculative workshops with designers and architects. These reveal fundamental tensions between proactive assistance and user autonomy, and between personalization and public ownership.

    \item An empirical characterization of multimodal interaction with robotic surfaces, drawn from a Wizard-of-Oz elicitation study. This includes the development of a modal entropy metric, which quantifies the diversity and evolution of a user's interaction strategies as they discover system capabilities.

    \item A set of research directions for cyber-physical architecture, synthesizing results from both studies.
    
\end{enumerate}

Together, the two studies point to a somewhat paradoxical finding: users want to feel in control of their environment, yet — consciously or not — they expect the environment to infer context that they never explicitly articulate. Effective control techniques, therefore, may require granting robotic environments the very type of agency that users sometimes claim to distrust. Navigating this trade-off emerges as a central design challenge for the field.

Our work concludes with a proposal for reframing interaction with cyber-physical architecture: rather than mapping discrete user actions to system responses, our findings make the case for a dynamic model of user intent, that continuously integrates multimodal evidence. This reframing, we argue, could help address the tensions unveiled by our studies, and thus may be an essential step in achieving fluent interaction \cite{hoffman2019evaluating} with robotic environments.

\section{Related Work}

A shape-changing environment is, in many ways, a robot that you can walk inside of \cite{kumar2024design}. To develop effective techniques for interacting with these structures, we draw on perspectives from tangible computing, room-scale actuation, and human-robot collaboration.

\subsection{Foundations}

Our research builds on work by Hong et. al \cite{hong2023interacting}, who conducted one of the first (to our knowledge) ideation and elicitation studies on the subject of large-scale shape-changing interfaces. In their initial study, the authors hosted a 30-minute brainstorming session, where small groups of participants worked together to generate application ideas for a hypothetical, wall-sized shape display (resembling inFORM \cite{follmer2013inform}, but turned on its side). A recurring theme was the concept of reconfigurable spaces — participants imagined situations in which seating and tabletops could be constructed when needed, and removed when not in use. 

To further explore this theme, the authors then conducted an elicitation study that focused first and foremost on the concept of explicit reconfiguration. Using a virtual shape display (a series of images, projected onto a wall), the authors asked users to perform actions that corresponded to various geometric changes, such as scaling or rotating a pattern of actuated cells. By design, the methodology mimicked that of Wobbrock et al. \cite{wobbrock2005maximizing, wobbrock2009user} — users were presented with a “before” and “after” state, and were instructed to come up with an action that could have triggered that change.

Most often, the users enacted gestures (using one or both hands), although this varied depending upon proximity. Users that were at least one meter away from the wall, for instance, would occasionally abandon gestures in favor of tool use (e.g. drawing on a phone screen) or voice commands. When standing up close, however, they opted for more hands-on techniques — a majority (16 out of 20) said that they would want to touch the wall when tasked with transforming it from one fine-grained state to another.

Interestingly, we found that touch was \emph{not} the dominant mode in our own task-based elicitation study — we saw instead a diverse set of multimodal interaction techniques, even when the wall was well-within reach. This suggests that while direct manipulation is a powerful paradigm for explicit state transformation, the interaction strategies for more goal-oriented, collaborative tasks are more complex. Our work seeks to understand how users navigate these scenarios, and expands on the role of agency in shape-changing environments.

\subsection{Intelligent Environments}

A long-standing goal in computing (specifically, within HCI) has been the development of environments that are imbued with ambient intelligence \cite{cook2009ambient, dunne2021survey}. These are physical spaces which can sense and respond to human activity. While early research often prioritized functional optimization, these environments increasingly target a broader spectrum of human experiences, including emotional well-being, social connectedness, and democratic engagement \cite{rao2025we}.

Initially born out of research in ubiquitous computing \cite{weiser1991computer}, these environments have appeared in various forms throughout the last three decades. Early work explored interactive tabletops \cite{wellner1991digitaldesk, wellner1993interacting}, augmented whiteboards \cite{stafford1996brightboard}, and ambient ceiling displays \cite{ishii1998ambientroom} — connecting digital information to specific physical objects, often with tangible controls. This later expanded to room-scale projection mapping, with influential projects like IllumiRoom \cite{jones2013illumiroom} demonstrating highly immersive visual experiences. More recently, systems like Dynamicland \cite{Victor2024DynamiclandIntro, bachmann2021promise} have fully embraced the “room-as-a-computer” paradigm, with a particular emphasis on the human-to-human collaborative opportunities that these environments afford. 

Most of these systems use cameras as the primary input device, though researchers have also explored additional modes for room-scale interaction. Often, these leverage sensors that are embedded into architectural elements: electromagnetic grids \cite{zhang2018wall}, optical fibers \cite{swaminathan2020optistructures}, load cells \cite{yoshida2022flexel}, or conductive paint \cite{wessely2020sprayable}. Coupled with machine learning algorithms, these methods enable off-body activity recognition \cite{das2023explainable}, allowing environments to understand both where users are, and what users are doing.

While there has been great progress in sensors and displays, a recent (and ambitious) goal is to enable the environment to actuate — to change its physical form \cite{green2016architectural, vazquez2019shape}. In HCI, this has been explored through reconfigurable walls \cite{gonzalez2023constraint, balci2025manifesting, larrea2015arkits}, floors \cite{je2021elevate, teng2019tilepop}, and furniture \cite{hauser2020roombots, suzuki2020roomshift}. In galleries, building lobbies, and public spaces, these installations often appear under the name “kinetic walls” \cite{Washabaugh2019KineticBrickWall, HyperMatrix2012, rozin2001wooden}. On an even larger scale, actuated building facades like the Al Bahr Towers \cite{attia2017evaluation}, and reactive pavilions \cite{strobel2018lumen, Sung2011Bloom, Wicaksono2023LivingKnitworkPavilion} change state in response to environmental data or human presence — for instance, providing reactive sunshading.

\subsection{Multimodal Interaction}

As environments become more capable, the channels of communication must become richer. Multimodal interaction, which combines inputs like voice, gesture, and touch, offers a more natural and expressive way for humans to communicate with machines. The seminal example is “Put-That-There” \cite{bolt1980put, schmandt1982intelligent}, which demonstrated how a deictic gesture (e.g. pointing) could resolve the ambiguity of a spoken command, and vice-versa.

This is especially important in shared physical environments, where achieving a joint understanding between humans and robots is a central goal \cite{chai2014collaborative}. Robots often use a combination of gaze, gesture, and speech to establish common ground and communicate intent \cite{huang2010joint}, allowing for the kind of fluid, back-and-forth exchange that Hoffman and Breazeal describe as “fluent” collaboration \cite{Hoffman2008Achieving}. For instance, when picking items from a shelf, a warehouse robot could concurrently parse vocal and gestural cues from a human partner, helping to retrieve specific objects in cluttered environments \cite{pathak2025multi}. Recently, advances in transformer architectures have greatly enhanced the perceptual capabilities of such robotic systems \cite{team2025gemini}. However, even state-of-the-art models can still can struggle over longer time horizons \cite{ayub2025continual}.

In addition to perceiving human input, embodied agents can also use their physicality to enhance communication \cite{Sauppe2014diectics}.  Compared to voice-only interfaces, these agents can be more engaging, and users have demonstrated higher situational awareness when interacting with them \cite{luria2017comparing}. These developments point towards richer possibilities that speculative design can help us explore.

\subsection{Speculative Design}

Speculative design is a research practice that uses sketches, objects, and fictions to explore possible futures (as opposed to near-term products). Its primary purpose is to challenge assumptions and provoke debate about the social, cultural, and ethical implications of emerging technologies \cite{malpass2013between}. The result is a shift in thinking from what is \emph{probable} to what is \emph{preferable}, which can help guide the design of human-interfacing systems \cite{candy2010futures}.

The field makes a distinction between affirmative design, which reinforces prevailing market logics, and critical design, which uses a form of “social dreaming” to prompt reflection \cite{dunne2024speculative}. A complementary approach is design fiction \cite{sterling2005shaping}, which employs diegetic prototypes — imaginary products that exist within a film or narrative  — to test and explore ideas before they are actually built.

Increasingly, these methods have been adopted within HCI — which has in turn spurred the development of hybrid techniques that combine speculation and elicitation. Through “speed dating”, for instance, groups of participants are exposed to a rapid showcase of future scenarios, prompting reactions that are noted by the investigators \cite{zimmerman2017speed}. Other methods involve theatrical performance, immersing users in potential scenarios while observing their responses \cite{lindley2015back}.

This is particularly relevant for human-robot interaction, aspects of which are, arguably, inherently speculative \cite{winkle2025robots}. Certain robotic form factors may be “maladapted” to non-industrial contexts \cite{auger2014living}, and require reconceptualization for domestic settings. Speculative methods are one way to attack this, engaging stakeholders through role-playing \cite{Katharina2024approaching}, brainstorming \cite{sturdee2015public}, and focused discussions \cite{grafstrom2022speculative}. For shape-changing interfaces in particular, design fictions can be used to address the central question of practical utility \cite{rasmussen2012shape}, visualizing how shape-change might manifest in everyday life \cite{sturdee2017using}. Participants in our workshops explored this through sketching and scenario-building, which can be a powerful method of inquiry \cite{sturdee2019sketching}.

\section{Design Workshops}

A central step for any new technology is articulating its value. When we introduce cyber-physical architecture, what actual problems do we solve, and what new experiences do we enable? To answer this, we conducted four speculative design workshops, aimed at uncovering characteristics that could pull shape-changing environments towards real-world adoption. Beyond just cataloging applications (the what), our study attempts to reveal the core values that underlie them (the why).

\begin{figure}[t]
  \centering
  \includegraphics[width=8.45cm]{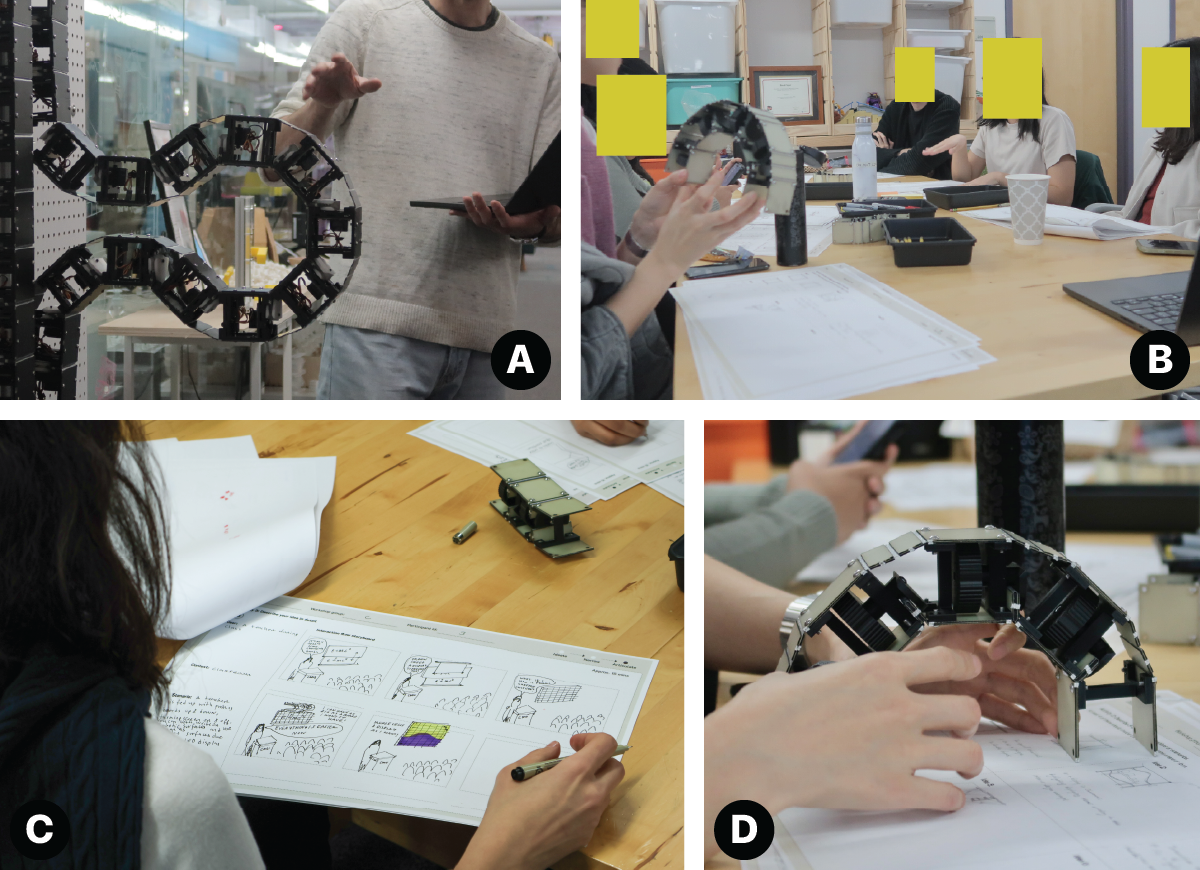}
  \caption{Participants in our design workshops were able to observe our robotic wall (a), and could request shape-changing demonstrations from our facilitator. They were also given a toy model to explore (d), which could be manually reconfigured. Throughout the workshop, participants sketched scenarios (c) and discussed hopes and concerns (d) related to cyber-physical architecture.}
  \Description{Photographs from a design workshop showing participants engaging with a shape-changing robotic wall, sketching scenarios on paper, and exploring a small reconfigurable toy model of the mechanism.}
\end{figure}

\subsection{Participants and Procedures}

For our initial exploration, we wanted to construct a participant pool that had been trained to think critically about spatial relationships and human needs. We therefore recruited 20 participants, explicitly filtering for those with experience in either architecture or design. Selecting one specific demographic (e.g. older adults, wheelchair users, etc.) would have been premature at this stage. 

Our participants had varying degrees of industry experience (ranging from 2 to 11 years), and were between 24 and 34 years old (median age: 27). At the time of the workshop, all were enrolled in either a Masters or PhD program.

We split these participants into four groups (4-6 participants each), and held a separate, 3-hour workshop for each group. This split was based primarily on participant availability, though we also endeavored to keep an even distribution of ages and backgrounds across the four sessions.

Each session was conducted in the same manner. We began by introducing the participants to our robotic wall, demonstrating basic transformations (simple curved forms such as the one in Figure 3a) and briefly explaining the principle of operation. We also gave each of them a small, non-actuated toy model to experiment with (Figure 3d). We did not discuss any example use cases, as we didn’t want to bias the participants towards one particular context or set of applications. This introduction lasted 20 minutes.

Immediately afterwards, we led participants through a set of three design activities. Though each activity was to be completed independently, we did not prohibit participants from talking with each other, and some short, spontaneous conversations did occur.

\begin{figure*}[t]
  \centering
  \includegraphics[width=\textwidth]{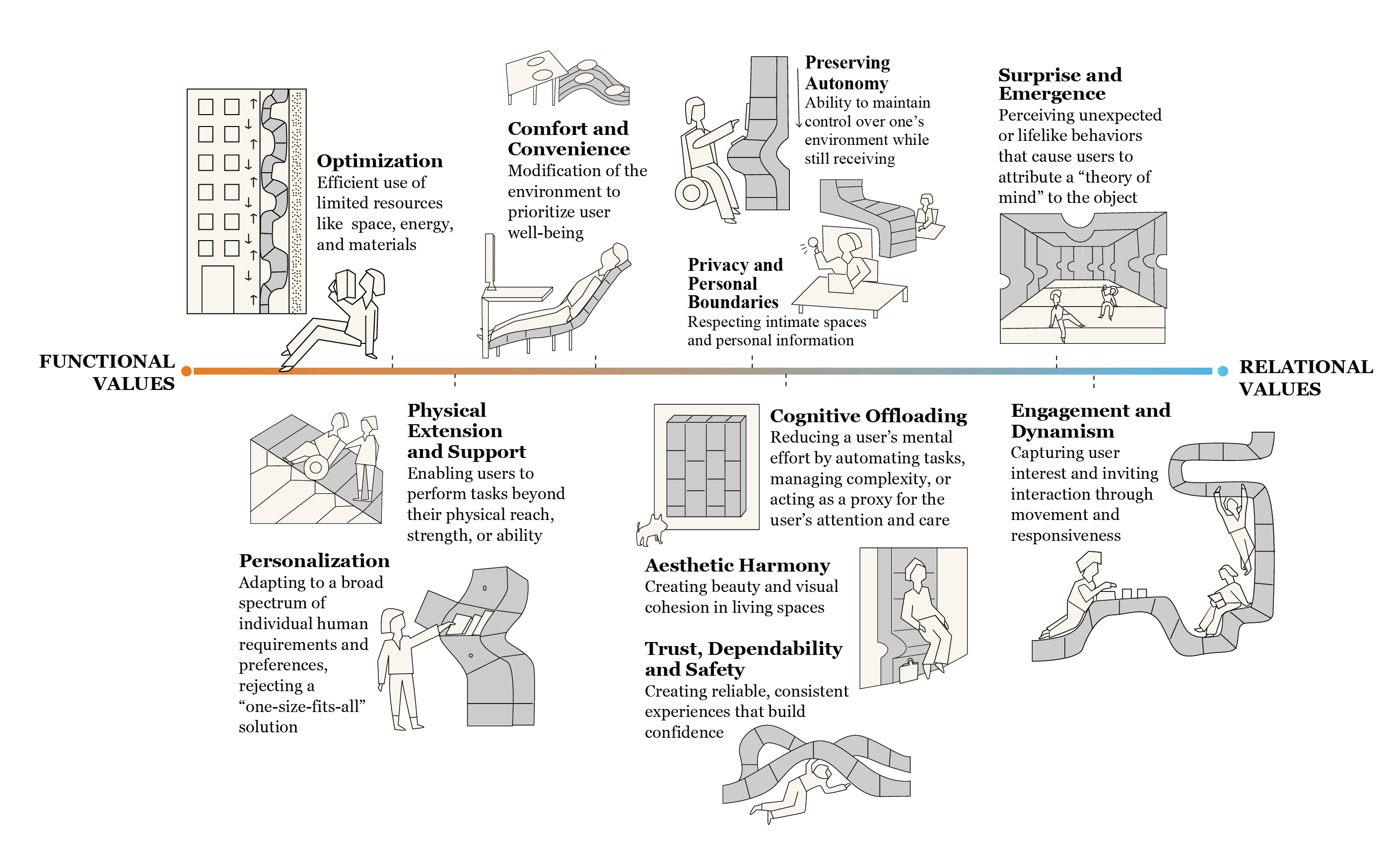}
    \caption{Analyzing the data from our design workshops, we identified eleven values which underpinned the scenarios our participants created. Some values are best categorized as "functional" (associated with tangible and pragmatic concerns), while others are best categorized as "relational" (describing psychological qualities of human-machine interaction). During our workshop discussions, tensions emerged between many of these values.}
  \Description{Diagram mapping eleven design values along a spectrum from functional to relational. Functional values include optimization, comfort and convenience, privacy and personal boundaries, physical extension and support, personalization, cognitive offloading, aesthetic harmony, and trust, dependability and safety. Relational values include preserving autonomy, surprise and emergence, and engagement and dynamism. Each value is defined and shown with a small illustrative sketch.}
\end{figure*}

The three activities were:

\begin{enumerate}
    \item \textbf{IDEATION.} For 25 minutes, participants were instructed to list, sketch, or doodle any concepts that involved a robotic surface. There were no explicit constraints, though we did ask them to consider the questions: “Where do you see these surfaces in the future?”, and “How might these surfaces become integrated into everyday life?”.
    
    \item \textbf{NARROWING.} After brainstorming, participants had 15 minutes to choose their favorite ideas and add additional detail, outlining specific applications or scenarios. We asked them to describe six ideas, with the stipulation that at least four of them should feature our specific style of robotic wall (as opposed to a pin display \cite{follmer2013inform} or similar).
    
    \item \textbf{ARTICULATION.} Finally, we gave participants 60 minutes to construct interaction flows (in the form of storyboards) for four of their ideas. We also asked them to clearly specify the context and scenario associated with each idea.
\end{enumerate}

Following these activities, we moderated an hour-long group discussion where participants presented their interaction flows (Fig 2b). We steered the conversation by asking a few key questions: How should agency be shared between users and walls? What tangible qualities of the robotic surface led the participants to their ideas? How would they extend their ideas if there were no technological constraints? And finally, what were the underlying values and concerns that motivated these scenarios? 

\subsection{Analysis Method}

The technique of qualitative content analysis \cite{schreier2012qualitative} allows for the systematic examination of communication and interaction. Since our study explored design-oriented data (rather than lived experiences), we found this method appropriate.

We began with data familiarization, examining all storyboards and discussion transcripts to gain a comprehensive understanding of the material. From this initial review, we identified seven overarching topics that were discussed in the workshops: form, function, values, contexts, users, ownership, and interaction techniques.

We then proceeded to code the transcripts and storyboards: identifying individual sketches, scenarios, and conversations that belonged to each of the seven topics. Within these topics, we identified more detailed codes. For example, the “context” topic included specific settings such as “home”, “outdoors”, “café”, and “office”. The “ownership” topic included subcodes such as “private”, “public”, and “personal”.

Next, we extracted all quotations in the “values” topic, and categorized them through a iterative coding process. Refining these categories, we arrived at a set of eleven values that underlay the storyboarded scenarios.

\subsection{Findings}

Is there real value in a shape-changing environment? These design workshops were born out of an initial concern — that despite the technological allure of “robotic walls”, the actual applications proposed in prior work can sometimes appear a bit contrived. We hypothesized that this could be due to (1) insufficient exploration of key contexts, and (2) limitations of the technology in its current instantiation (i.e. perhaps there exists a more appropriate form factor or interaction mode). Through these workshops, we aimed to identify values that could guide interested researchers towards building “the right thing” \cite{jung2025making}.

Eleven values emerged during our workshop analysis, which we arranged on a spectrum from “most functional” to “most relational”. Functional values encompass the pragmatic and performance-oriented qualities of shape-changing technology. Relational values capture interactions and dependencies between people and a robotic environment. We encourage the reader to examine Figure 4 for an overview of the values, before continuing to the more detailed narrative in the following subsections.

\begin{figure}[t]
  \centering
  \includegraphics[width=8.45cm]{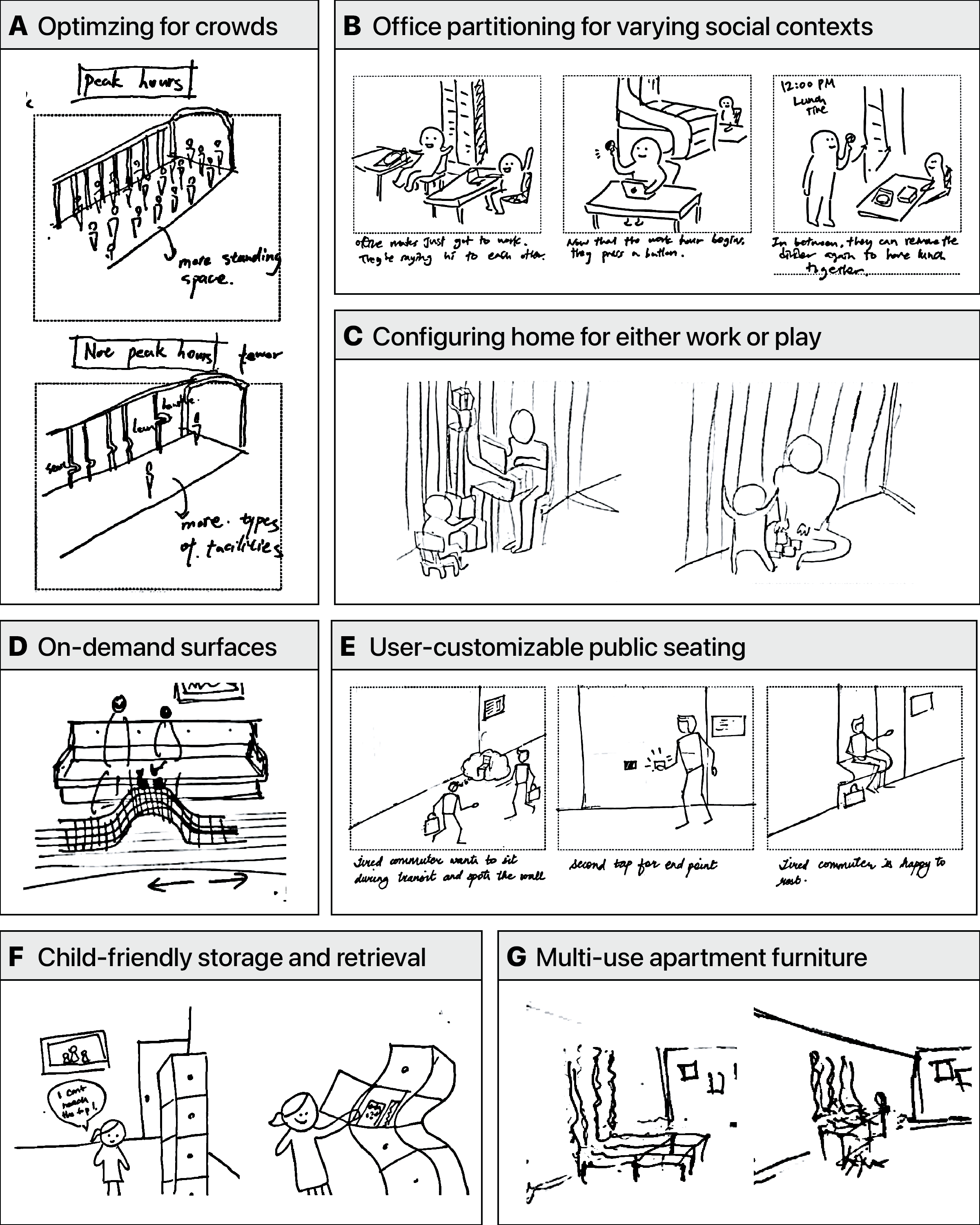}
  \caption{Participants developed storyboard scenarios illustrating the value of space optimization. The examples include a) managing peak and off-peak crowding, b) partitioning spaces for different social interactions, c) configuring homes for work and leisure, d) creating on-demand surfaces, e) customizing public seating, f) enabling child-friendly storage, and g) designing multi-use apartment furniture.}
  \Description{The figure contains seven storyboard sketches created by participants to illustrate value of space optimization. The panels show concepts such as adjusting layouts for crowded and uncrowded times, adding partitions for social contexts, reconfiguring homes for work or play, surfaces that change on demand, seating that can be customized in public settings, storage designed for children’s use, and multi-use apartment furniture.}
\end{figure}

\subsubsection{Resource Optimization}

Noting the inherent reconfigurability of the wall, many participants suggested applications that prioritized multifunctionality and reuse. In dense urban environments, where living space is limited, they imagined dynamic room partitions, expandable storage, and on-demand seating in small apartments and offices (Figure 5). “[People] are looking for solutions where they can sort of optimize their space,” mentioned one participant, drawing from their experience as an architect. Another agreed: “The most important thing [clients] say is more space and less cost. That's how the ratio goes.”

This mode of space optimization was proposed for public areas as well: namely, train cars, subway stations, and airports (Figure 5a, 5e). “You don’t necessarily need permanent furniture,” shared one participant, “but if it’s a long transit, I just want to sit and rest for a bit.”

At larger scales, many participants envisioned the technology as an architectural element for energy savings, serving as a temperature-sensitive facade that could adjust to block solar heat and reduce cooling demands (Figure 6). This was often described as occurring automatically, though some participants sketched systems for manual control (e.g. wall-mounted sliders, mobile dashboards).

Shape-changing structures were also seen as a way to conserve and reuse materials. Participants described desks and chairs that could “grow up together with [a] child”, as well as building blocks that could be reconfigured over time. “After two months, [maybe] he doesn’t want his books, he wants a [music] collection,” said one participant, describing an imagined user. “[The wall] is multi-use… it doesn’t necessarily have to be treated as a bookshelf.”

But there was also a tension among participants, centered around the large upfront ecological cost of the proposed “resource-saving” applications. How many pieces of furniture does the wall need to replace, for instance, in order to justify the initial investment? Participants conceded that while small living spaces may be a promising context, reconfigurability alone may not provide sufficient value. “Even then, people get very inventive,” one participant argued. “So… for it to have to be a robot wall, I feel like there’s [got to be] some sort of automation.”

\begin{figure}[t]
  \centering
  \includegraphics[width=8.45cm]{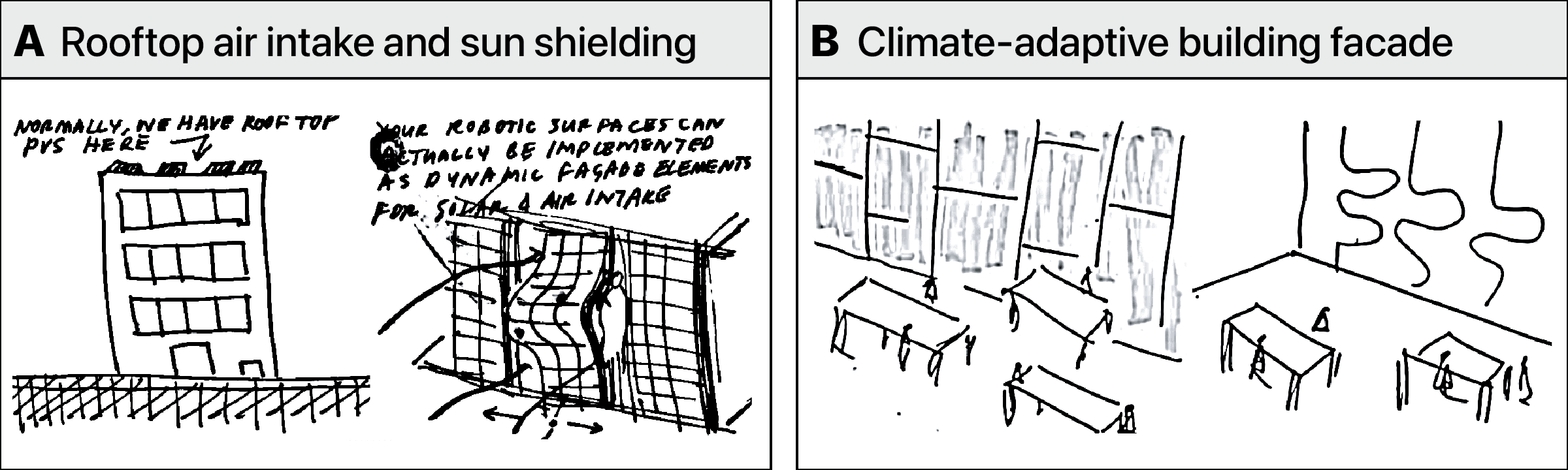}
  \caption{Storyboard scenarios illustrate architectural-scale applications for energy efficiency: a) rooftop air-intake and sun-shielding structures, and b) climate-adaptive facades that adjust to block solar heat and reduce cooling demand.}
  \Description{The figure contains two storyboard sketches. The first shows a building with rooftop structures designed for air intake and shading. The second shows an interior view of a building with a facade that changes shape in response to climate, depicted as adjustable wall and window elements that reduce solar heat.}
\end{figure}

\subsubsection{Cognitive Offloading}

Many participants saw value in offloading mental tasks to the environment — automating minor functions, or permitting the wall to move proactively. “The usefulness of it being a robot is [that] I don’t have to do anything,” declared one participant. “It just does it.” In the minds of some participants, these more agentic behaviors seemed to warrant the complexity of a technological solution.

Several scenarios involved “just-in-time” support, in which the wall would handle finer details so that users could remain focused on a primary activity. One participant sketched a kitchen backsplash that served as an assistant chef: delivering ingredients, removing clutter, and holding a mobile phone to display recipe steps (Figure 7c). Another described a chair-like form that offered posture support over long work sessions, and could prompt the user to take a break and stretch.

Other scenarios ranged from simple caregiving, such as playing with pets or adjusting sunlight for plants, to more critical support, such as detecting a fall for an elderly user. There was, of course, a caveat: the automation had to be competent. The promise of assistance raises the bar for functional utility \cite{ray2008people},  and participants warned that a slow or clumsy robot would create more frustration than it solved. If a command took too long to execute, one participant said that they would simply give up: “okay, whatever, I will do it.” As is the case in other robotic systems, successful offloading relies heavily on established trust \cite{chi2023people}, which participants noted can be fragile.

\begin{figure}[t]
  \centering
  \includegraphics[width=8.45cm]{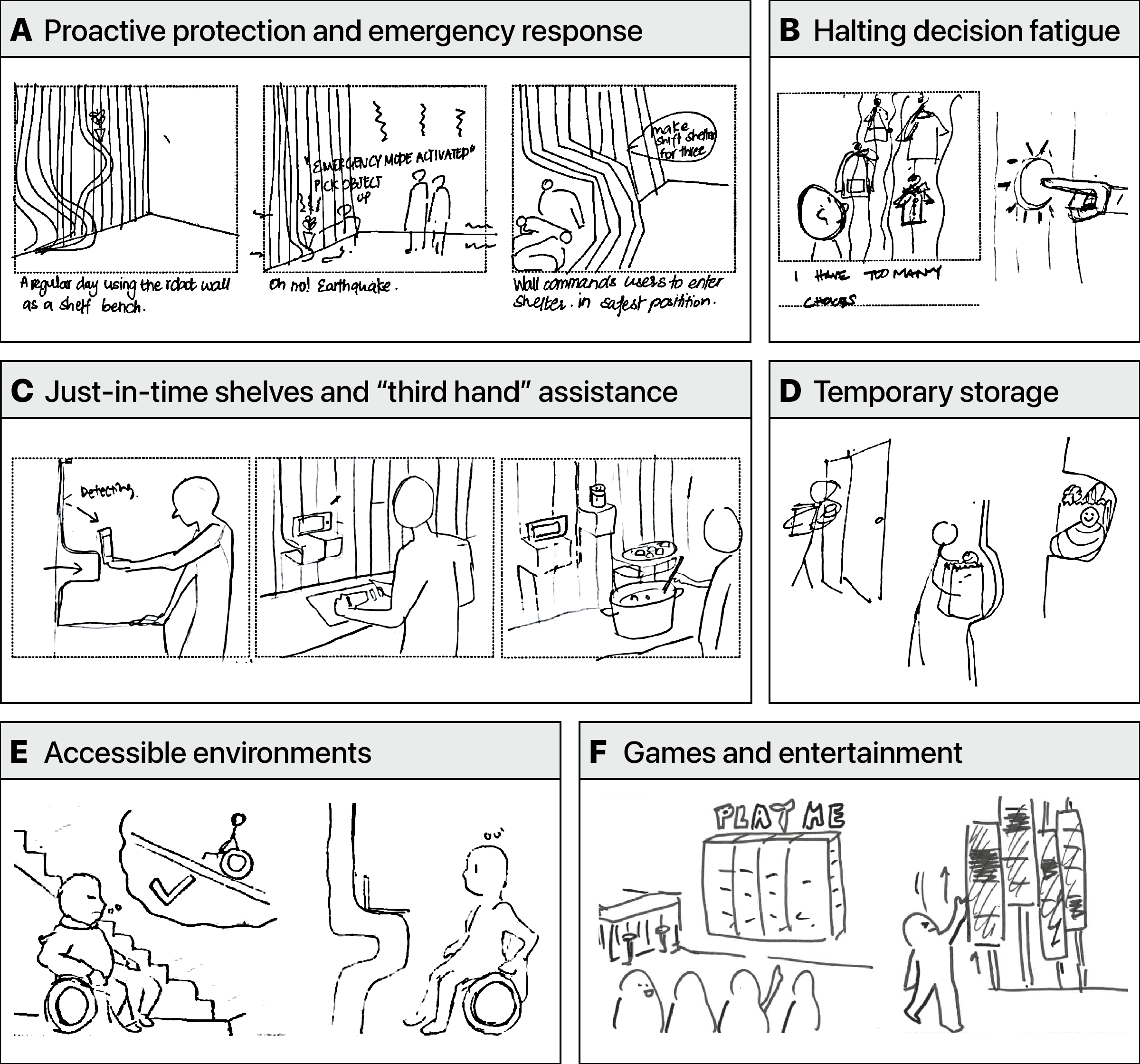}
  \caption{Storyboard scenarios depict protective and assistive applications of responsive surfaces. Examples include a) transforming into emergency shelters, b) reducing decision fatigue by simplifying choices, c) providing just-in-time shelving and assistance in the kitchen, d) enabling temporary storage, e) supporting accessible environments, and f) applications with games and entertainment.}
  \Description{The figure contains six storyboard sketches. One shows walls reconfiguring into protective shelters during emergencies. Another depicts surfaces helping reduce decision fatigue by narrowing clothing options. A kitchen backsplash is drawn as a helper that provides shelves, delivers ingredients, and holds a phone. One panel shows temporary wall storage for bags. Another illustrates surfaces creating accessible environments for wheelchair users and adaptable seating. The final sketch shows a wall transforming into an interactive game for entertainment.}
\end{figure}

\subsubsection{Preserving Autonomy}

Sometimes at odds with the desire for proactive assistance, participants emphasized that any automation must still respect user autonomy. Multiple scenarios reflected a preference for explicit, user-initiated interactions. “For me, it's mostly on the human to start,” said one participant. “You hit a button and it does something.” (Figure 7b).

This desire for control was often rooted in a practical concern about the risks of misinterpretation. “I do not want it to predict what I want, because what if it predicts something wrong?” asked one participant. “I live on the third floor,” joked another, “What if it throws me out the window?”. To give users reliable agency, participants proposed graphical user interfaces for precise control, and simple voice commands for more atomized interaction. Automation was more readily accepted in cases where the functions were temporary and reversible (i.e. the costs of misinterpretation were lower).

A related discussion emerged around the degree of “surveillance” that would be necessary to facilitate proactive robotic behavior. A few participants indicated that they would be more amenable to sensing mechanisms if the corresponding behavior resembled a physical reflex, as opposed to a more calculated decision. “It's not making inferences about you as a person,” said one participant, “It’s only making an inference about where your body is going.”

\subsubsection{Privacy and Personal Boundaries}

Conversations about agency and proactivity led participants to consider the privacy implications that come along with a robotic environment. Interestingly, this was discussed in two modes: one in which the wall was portrayed as a prying, unwanted intruder; and another in which the wall took an active, protective role in establishing boundaries between a user and other people.

In the more protective case, participants described office cubicles that could promote either focused work or social interaction; and apartment partitions that could move to obscure sensitive areas (Figure 6b). “If you’re staying in a studio apartment and your friend comes to visit you, you might not want him to see your sleeping space,” said one participant, “because that’s more personal to you.”

Participants sketched “secret” spaces: camouflaged doors with fingerprint access, and concealed storage nooks in bathrooms and bedrooms. “You can hide little objects,” suggested a participant. One amusing scenario involved a couple  using a shape-changing bed as a more “active” enforcer of personal boundaries: “If your husband’s irritating you, just press a button… you’re gone off the bed.”

The second discussion mode encompassed perceptual and data privacy. We noticed mixed views on sensing and predictive features. Some participants were more accepting: “If it really gives enough value for me, then I would [give] out my right to be seen.”  Others, however, preferred manual control over predictive actions, worried about misinterpretation and unwanted outcomes. There was some inherent vulnerability in sharing space with an intelligent environment.

\subsubsection{Trust, Dependability, and Safety}

As participants crafted increasingly sophisticated robotic behaviors, they also raised their expectations for trust and reliability.

Participants envisioned trust building gradually, almost like training a pet. “Anything new is always jarring,” reflected one participant, “it takes time to get used to [it].” Starting with low expectations seemed important, allowing new capabilities to be introduced bit by bit. Legible behavior was also crucial. “If it's approaching me and it naturally shows me that it's slowing down, [then] I can trust it,” explained a participant.

Yet the margin for error was small. “If it ever knocks something out of place, even the tiniest thing,” warned one participant, “[then] I can't trust it after that point, right?” This became especially critical for vulnerable populations. For some scenarios that the participants described — wheelchair transfer assistance, rehabilitation support — a system failure could cause serious personal harm.

Physical form also influenced trust perceptions in ways that transcended pure functionality. “The softness of it... reduced my worry," shared one participant, referring to the “rounded” edges of our shelf configuration (Figure 1a).  Tactile qualities and visual cues could be important in establishing user confidence.

Participants also imagined these surfaces in “protective” roles: transforming into emergency shelters during earthquakes, lifting precious objects during floods, and serving as fireproof partitions (Figure 7a). While these applications highlighted the demanding standards for trustworthiness, they also revealed participants’ excitement about the wall's physical capabilities.

\begin{figure}[t]
  \centering
  \includegraphics[width=8.45cm]{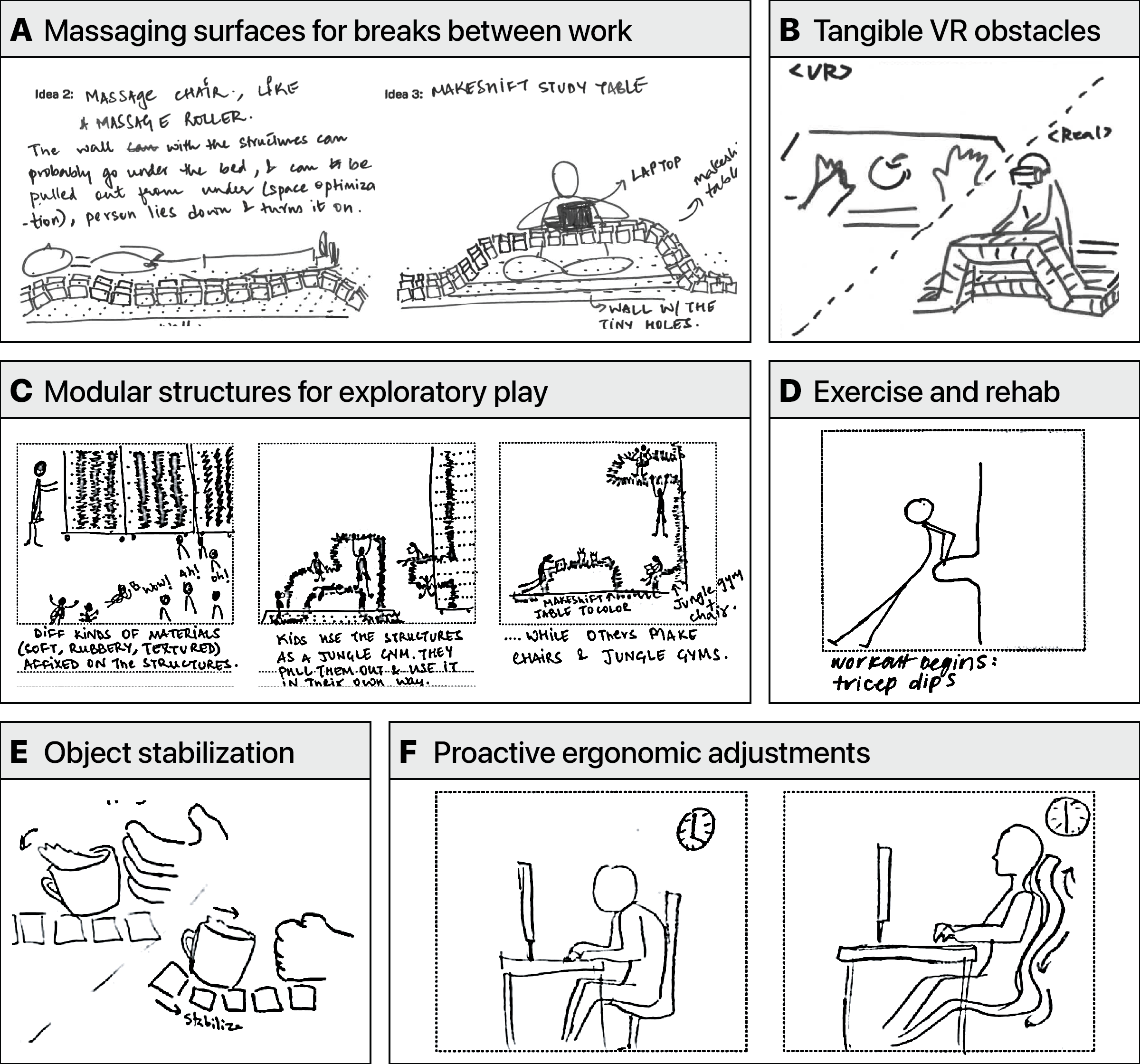}
  \caption{Storyboard scenarios illustrate diverse applications of responsive surfaces. Examples include a) massaging surfaces for breaks during work, b) tangible VR obstacles, c) modular structures for exploratory play, d) exercise and rehabilitation support, e) stabilizing objects to prevent spills, and f) ergonomic adjustments in office settings.}
  \Description{The figure contains six storyboard sketches. The first shows surfaces reconfigured into a massage bed and resting area during work breaks. The second depicts a person navigating obstacles in a virtual reality setting. The third shows children climbing and playing on modular structures. The fourth illustrates a person performing exercise for rehabilitation. The fifth shows a table stabilizing cups to prevent spills. The sixth depicts an office chair reshaping to improve posture while a person works at a desk.}
\end{figure}

\subsubsection{Physical Extension and Support}

Robotic surfaces were often proposed as tools for physical augmentation — extending human reach, strength, and precision beyond natural limitations.

Many participants gravitated toward industrial applications for worker safety, sketching surfaces that could serve as dynamic scaffolding, or could transport construction materials. “Sometimes you need to carry things [between] floors before the lift is made,” explained one participant.

A robot that is powerful enough for “heavy-duty” tasks, however, also carries a greater risk of harm. This pushed participants toward form factors quite different from those on factory floors. There was a  large emphasis on collapsible structures, lightweight panels, and hand-configurable modules. Many drew clear inspiration from our prototype, though in practice, these large-scale applications would likely require a more robust constraint mechanism.

Participants also envisioned scenarios where the environment could act as a “helping hand” during cognitively demanding tasks. In a hospital, shape-changing lights in an operating room could reposition themselves as a surgery progressed. In an office, a shape-changing table could stabilize objects that were accidentally knocked over, preventing damage during delicate work (Figure 8e).

Some of the most compelling applications emerged in therapeutic contexts, where the wall could provide both physical support and procedural guidance. “Consider a physical therapy approach,” said one participant. “If someone doesn't remember how to do a particular exercise, they can lean up against the wall, and the wall comes out to meet them in the place where they should be.” Some participants even imagined the wall verbally speaking to users, to guide them through weight training and rehabilitative activities (Figure 8d).

Accessibility emerged as a particularly rich area for exploration, with participants describing surfaces that could adapt to different physical needs. They sketched walls that could bring objects down from high places, actuated handrails that could help people stand up, and load-bearing surfaces that could transform between stairs and ramps. Rather than designing for an “average” user, participants envisioned environments that could meet people where they are. “I see that as an opportunity,” said one.

\subsubsection{Personalization}

Personalization seemed straightforward to participants: it would be beneficial to have surfaces that could tailor themselves to individual bodies, preferences, and needs. But as discussions deepened, it became apparent that this value was tied to the question of “who controls the space”.

In personally-owned spaces, participants embraced elaborate customization scenarios. They described walls that could shift from “wavy” to “straight lines” based on mood, surfaces that responded to musical beats, and environments that adapted as children grew or lifestyles changed. “Different individuals have different body types,” reflected one participant, “and [we] can customize it to the needs of everyone.”

This was especially important for accessibility, particularly within family households. Rather than requiring permanent modifications that might stigmatize some users or inconvenience others, participants envisioned surfaces that could temporarily accommodate wheelchairs, then reconfigure for walking family members. The same kitchen counter could lower for a child learning to cook, adjust for a parent using mobility aids, then return to standard height for daily use.

In privately-controlled spaces — offices, stores, gyms, theaters — personalization introduced competing interests. Though participants were enthusiastic about adjustable counters in coffee shops, and walls in climbing gyms that adapted to different skill levels, they also recognized that businesses must balance individual customization with broader operational needs.

The most provocative tensions surfaced around public spaces. While participants initially described benches that could adjust for different users, the implications quickly became uncomfortable. “I think it gives a lot of power over public spaces to people who can afford this kind of technology,” warned one participant. “You can very easily make that anti-homeless architecture, right? By just having every public bench be [made of] this, and then at night they go down.”

This concern extended to the visual character of shared environments as well. When every surface can transform at will, who determines what the neighborhood should look like?

\subsubsection{Aesthetic Harmony}

For many participants, the most compelling aesthetic quality was the technology’s ability to disappear. Having the capacity to “move and create forms… and then go back to nothing,” was the “unique selling proposition,” as one participant put it.

This visual cohesion was also pitched as a privacy-preserving feature. “I thought of a concealed door,” shared one participant. “[This] part kind of curves into a push bar… [then] the door shuts and it again flattens out.”

Participants also described scenarios where instead of adding mechanical agents to a space (e.g. a robotic humanoid maid), the technology could become one with the space itself. They sketched undulating surfaces that could shuffle objects from room to room, moving stray items out of sight.

Overall, participants favored soft, curved forms over angular mechanical aesthetics, and envisioned context-specific finishes — wood for homes, stainless steel for kitchens, concrete for public spaces. Scale became another aesthetic dimension. Participants sketched small interior panels as well as entire building facades — which could create dynamic patterns and draw attention from passers-by on the street.

\subsubsection{Engagement and Dynamism}

Participants were drawn to the wall’s dynamic qualities, envisioning new modes of interaction that ranged from direct, tactile manipulation to the passive appreciation of ambient motion. 

The most hands-on scenarios involved a form of co-physical interaction, where users could manually shape their environment. One of the most vivid concepts was a large, child-friendly play structure that children could literally pull apart and reshape. “I was imagining this to be, like, a techie version of [Cas Holman's construction sets],” one participant explained, referencing a collection of large wooden toys for unstructured play (Figure 8c). This style of interaction would likely require some means of inferring intent. The robotic structure must be intelligent enough to know when to yield control to a child's push or pull, and when to stiffen to securely hold a new form.

Participants also envisioned these surfaces facilitating new forms of entertainment. Several imagined the wall giving physical form to virtual reality experiences, creating tangible obstacles that mirrored the digital world (Figure 8b). Others proposed more playful, arcade-style applications, such as a large-scale version of “whack-a-mole” (Figure 7f).

The surface’s capacity for more “freeform” motion proved equally compelling, and was further heightened by the concept of a concealed actuation mechanism. “If I'm not seeing what's inside, it's mystical,” explained one participant. “I don't know how this is happening.”

At a public scale, this dynamism could be a tool for civic and commercial appeal. Perhaps inspired by kinetic artworks, participants imagined building facades that could become local landmarks through their eye-catching transformations. “If you see something very interesting going on on a facade,” said one participant, “it automatically [brings] business.” These conversations hinted at a deeper interest in more spontaneous behavior.

\subsubsection{Surprise and Emergence}

Although dependability is often a requirement for developing trust, a compelling counterargument unfolded around the merits of unpredictability. Participants explored how emergent, surprising behaviors could encourage a user to attribute a theory of mind to the wall, viewing it less as a tool and more as a “living organism” with its own agency. Zooomorphic traits, like subtle breathing motions or responsive “nods” could give the sense of “being seen” \cite{zuckerman2020companionship} by the environment. This was not limited to ambient actions. One participant suggested that if a user's command was followed \emph{slightly} imprecisely, it could be interpreted as the wall balancing the user's instruction with its own internal priorities. They compared the relationship to having a pet dog.

A few participants associated surprise with intuition, describing scenarios in which uncertain user intentions might lead the wall to improvise a response. One participant claimed that user desires are often under-specified, and said that under those conditions, they would be more forgiving of rogue actions by the wall. “My mom gives me very vague instructions,” they empathized, “I need cues too.”

Perceived agency could also spur and reward curiosity. For instance, if a user observes the wall reacting to changes in musical tempo, they might reason that “it could react to [rhythmic] motion as well.” A surprising interaction might reveal one function, which in turn encourages the organic discovery of related capabilities.

\subsubsection{Comfort and Convenience}

Participants also saw value in an environment capable of providing proactive care. This included somatic comfort — massaging surfaces that could respond to strain (Figure 8a), or ergonomic seating that could shift forms after prolonged use (Figure 8f). It also included affective support, with proposals for environments that could regulate mood: a “calming room” with wave-like walls, and an energizing one with rapid motions and lighting.

Other participants described a form of logistical care, aimed at reducing the friction of daily life. In public, this included on-demand convenience, like temporary seating in transit hubs or retail shelves that could lower items on command. In the home, they pictured a domestic assistant: a wall that could “get [a] carrot for me” while a user cooks, or help carry meals from the kitchen to the dining room.

\subsection{Reconfigurability is Not Enough}

Our workshops confirmed that functional utility remains paramount. On the whole, gravitated toward scenarios where shape-changing surfaces performed tangible work —optimizing square footage, retrieving objects, supporting physical rehabilitation — rather than serving as ambient companions. This aligns with broader findings in robotics research, where users consistently prioritize practical assistance over social presence\cite{ray2008people}.

Yet shape-changing environments occupy a distinct category. Unlike a vacuum robot or warehouse automaton that disappears to complete a task, these structures remain continuously present, responding to occupants in real time. The relational dimension cannot be sidestepped: users must coordinate with them, interpret their movements, and negotiate shared space. Several participants also noted that the perceived expense of such installations would raise their expectations accordingly. An actuated wall that can not react to users intelligently might feel like an over-engineered waste of resources, particularly when some of largest purported benefits are space optimization and material reuse.

This observation motivated our second study. If shape-changing environments must respond with a degree of contextual awareness, then we first need to understand how users naturally attempt to communicate with them. What cues do they offer, and which do they leave implicit? The following elicitation study addresses these questions, examining the multimodal strategies that emerge when participants pursue concrete goals with a robotic wall.

\section{Elicitation Study}

Prior to this work, we, the authors, had our own set of assumptions for how users ought to control a robotic surface. The designers and architects in our speculative workshop had opinions on this as well — some of which ran counter to our own. To gather more concrete evidence regarding natural interaction techniques, we conducted an elicitation study using our prototype robotic wall (Figure 9). Our goal was not only to observe users’ initial actions, but also to understand how they react and adapt when the system makes mistakes.

This requires a slightly different approach to elicitation. Typically, HCI elicitation studies are conducted by presenting users with a set of “referents” — images, mock-ups, or animations that show an object or interface transitioning between states \cite{wobbrock2009user}. Participants are then asked to perform the action that they would have used in order to trigger that state transition.

Our approach draws instead from the “live” Wizard-of-Oz techniques more commonly seen in Human-Robot Interaction literature \cite{riek2012wizard}. Rather than showing participants a referent, we gave them a physical task to perform with the robotic wall. As they attempt the task, a human operator observes the scene and remotely generates responses. Though it requires more upfront setup, this methodology offers two key advantages. First, it allows us to naturally introduce system errors, and observe how the participants recover. Second, by focusing on a functional goal (as opposed to a singular state transition), we can elicit multi-step interaction sequences, which may unveil latent behavioral patterns.

\begin{figure}[t]
  \centering
  \includegraphics[width=8.45cm]{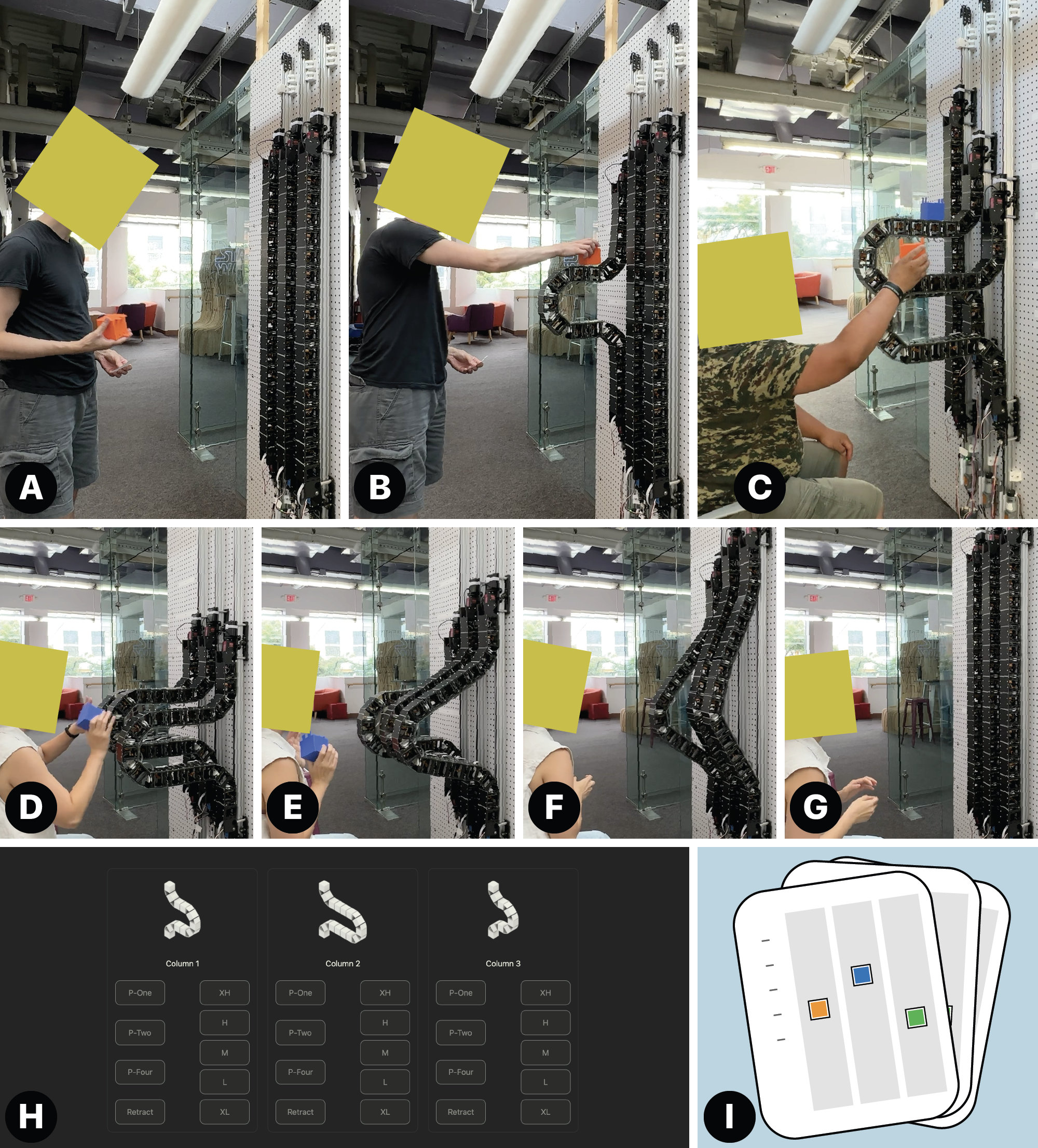}
  \caption{In our Wizard-of-Oz elicitation study, participants interacted with our shape-changing robotic wall. They were given a series of cards (i) and tasked with placing colored blocks on shelves at varying heights. Participants had to solicit shelves from the wall, both while standing (a,b) and sitting (c). After placing all blocks, and adjusting the heights of the shelves, they had to return the wall to the flat state (d-g). The robotic wall was controlled by human operator, using a graphical interface (h).}
  \Description{Photographs from an elicitation study showing participants interacting with a shape-changing robotic wall by placing colored blocks, observing the wall reconfigure into different forms, and using digital interfaces and diagrams to control or represent the wall’s behavior.}
\end{figure}

\subsection{Participants and Procedures}
We recruited 12 participants for our elicitation study (6 male, 6 female). They ranged in age from 18 to 38, with a median age of 25. Each one visited our lab independently, and each session lasted one hour.

When the participants entered, we first introduced them to our robotic wall (Figure 9). This was a modified version of the robotic surface presented in previous work \cite{gonzalez2023constraint}, now composed of three shape-changing columns that could move and reconfigure themselves in parallel. We showed participants a video of the wall in motion, but importantly, did not demonstrate any potential interaction techniques.

Each participant was given the same task, which was repeated throughout the study: to place a series of colored blocks at targeted positions on the wall (Figure 9a-c). To achieve this, participants would have to solicit a “shelf” from one or more columns, and adjust the height upwards or downwards if necessary. After all blocks were placed, participants would retrieve them, and then attempt to return the wall to the flat state (Figure 9d-g). For each trial, participants were given a randomized card which showed where the blocks should be placed (Figure 9i).

Participants were told that the wall could see them, hear them, and feel them. In reality, an experimenter was both overseeing and controlling the wall’s actions, using a remote graphical interface.

Since real-world systems may occasionally misinterpret user commands, we also probed users' recovery behaviors. At pseudo-random intervals, our interface suggested “errors” (small response delays, deployments at incorrect heights), which the experimenter would then incorporate into their performance. Interestingly, post-study interviews revealed that participants did not find these types of errors frustrating — perhaps because they were relatively low-impact and easily correctable (which would be consistent with prior work \cite{salem2015would}).

We selected this task for three reasons. First, it represents a simplified analog of many scenarios that arose during our design workshop — storage and retrieval, “dismissible” furniture, and personalized height adjustments. Second, the task is easily parameterized, enabling controlled variations (target heights, target columns) across trials. Third, because the task context is rather abstract, we anticipated that it would invite a range of interaction strategies, without biasing toward any single modality.

The study consisted of four stages, each with twelve block placements and retrievals. In the first and fourth stages, participants were free to use any interaction modality. In the second and third stages, we introduced a forced choice: a “voice only” mode and a “hands only” mode. Specifically, participants were told that certain sensing capabilities (i.e. audio, or vision/touch) had been deactivated. Through these restrictions, we aimed to elicit a wider range of techniques from participants that might have had a pre-existing bias towards one specific modality.

The ordering of these forced choice stages was randomized for each participant. Within each stage, half of the trials were conducted while the participant was standing up, and the other half were conducted while the participant was sitting in a chair (Figure 9c).

After all stages were completed, we conducted a ten-minute interview. We asked each participant to describe their preferences, frustrations, and recommendations; and then revealed the Wizard-of-Oz component of the study.

\subsection{Analysis Method}

We recorded all user sessions on video. We began our analysis by reviewing these recordings — developing a preliminary codebook by taking coarse notes on the voice, gesture, and touch actions that occurred.  Next, we conducted a second, fine-grained pass through the videos. In this pass, we logged timestamps and codes for each action, adding contextual details where necessary. Throughout the analysis, these codes were progressively refined.

This process resulted in over 2,000 identified actions, tagged with a set of roughly 200 individual codes. Afterwards, we grouped these codes hierarchically, ending with 50 primary codes and 194 total subcodes (across all three modalities). Unlike our design workshops, we chose \emph{not} to code the debriefing interviews, as these were relatively short and served primarily to clarify ambiguities that arose during the elicitation tasks.

\subsection{Overview of Findings}

There is no “one-size-fits-all” modality for interacting with robotic surfaces. Participants had varying preferences — some of which depended on the physical context (standing vs. sitting), the type of action (gross vs. fine-grained control), and the social environment (public vs. private). But perhaps most notably, many opted for a multimodal approach, fluidly mixing voice, gesture, and touch in order to accomplish the study tasks.

We can begin to visualize this data by plotting participants’ actions on a timeline, as shown in Figures 9 through 11. The plot at the bottom of Figure 10, for instance, shows the sequence of actions taken by Participant 12 throughout the course of the final stage. Note that for some points on the timeline, multiple actions were observed. For instance, a participant may have issued a vocal command (“make a shelf here”) along with an accompanying gesture (pointing at a desired location).

For each participant, we can also calculate a \emph{modal entropy} — a measure of the diversity of actions for each participant. Participant 2, for instance, who primarily interacted with the wall via a small set of touch actions, will have a lower modal entropy than Participant 5, who alternated between voice, gesture, and touch. These calculations are explained in Section 4.6, and are derived from the lower-dimensional action representations introduced in Sections 4.4 and 4.5.

By examining the modal entropies for the first and final stages of the study, we can create four qualitative groupings of participants. The groupings capture the participants’ preferred interaction style, as well as describe how these preferences changed over the course of the experiment (before and after exposure to the two “forced-choice” stages).

These groupings are displayed on the following page.

\pagebreak
\begin{figure}[t]
  \centering
  \includegraphics[width=8.45cm]{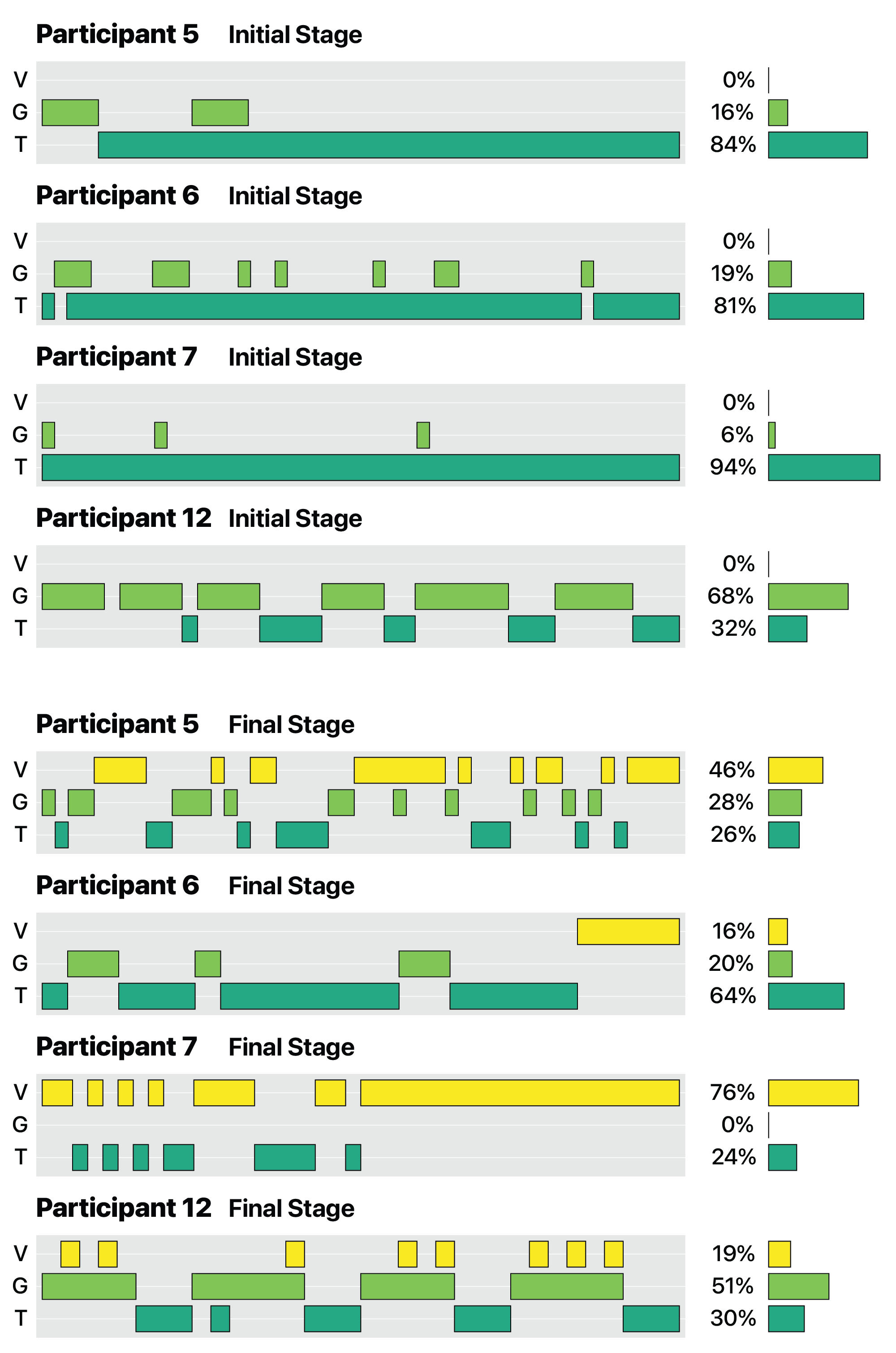}
  \caption{Timeline of vocalizations (V), gestures (G), and touch actions (T) for four participants (P5, P6, P7, P12), during both the initial and final stages of the study. }
  \Description{Timeline of vocalizations (V), gestures (G), and touch actions (T) for four participants (P5, P6, P7, P12), during both the initial and final stages of the study. These participants began the study favoring physical interactions, but by the final stage, had adopted more multimodal strategies. The right side of the figure displays the distribution of actions for each participant and stage.}
\end{figure}

\noindent \textbf{Group 1: Adopted Multi-Modal Preference.} The four participants in Figure 10 (P5, P6, P7, P12) began the study by using only physical interaction modes — gesture and touch. However, by the end of the study, they had incorporated voice commands into their repertoire. In the first stage, for instance, Participant 12 attempted to adjust the height of the shelves by holding an object at a desired location, waiting for the shelf to “meet” the object at the target spot. By the end of the study, they chose to first place the objects on the shelves (at any location), and then make smaller height adjustments with short voice commands (“go up”, “go down”). Similarly, Participant 5 initially treated the wall as something to be physically “molded” — interacting almost exclusively through touch. But in the final stage, they adopted a more diverse style, adding pointing gestures, verbal instructions (“move all the curves to the first marker”), and verbal confirmations (“okay”, “no”).   These shifts imply that there may be interaction techniques which, while not always readily apparent, become preferable once users  “discover” them. 

\pagebreak
\begin{figure}[t]
  \centering
  \includegraphics[width=8.45cm]{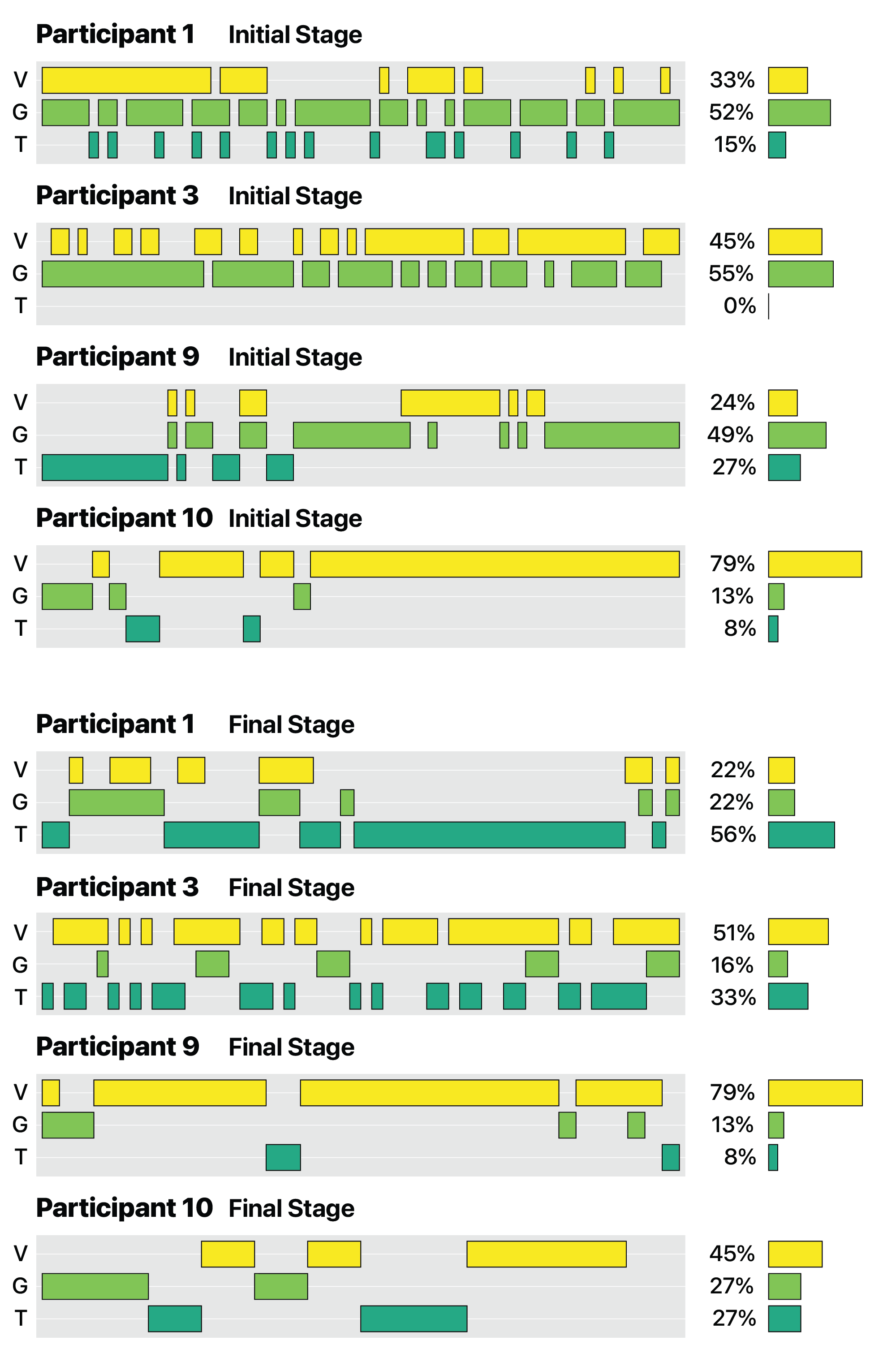}
  \caption{Timeline of vocalizations (V), gestures (G), and touch actions (T) for four participants (P1, P3, P9, P10), during both the initial and final stages of the study. These participants started and finished the study using multimodal strategies. The right side of the figure displays the distribution of actions for each participant and stage.}
  \Description{Timeline of vocalizations (V), gestures (G), and touch actions (T) for four participants (P1, P3, P9, P10), during both the initial and final stages of the study. These participants started and finished the study using multimodal strategies. The right side of the figure displays the distribution of actions for each participant and stage.}
\end{figure}

\noindent \textbf{Group 2: Stable Multi-Modal Preference.} Another third of the participants (P1, P3, P9, P10) used both vocal and physical interaction modes from the onset of the study (Figure 11). All retained this multimodal preference, though many altered their distribution of actions as the study progressed.  Participant 1, for example, increased their use of touch, particularly when soliciting new shelves from the flat wall. “I could just point to the [shelf] that I wanted to come out,” they said, “I feel it’s more direct for me.” For the same subtask, Participant 9 was drawn in the opposite direction, using voice commands to communicate intentions while concurrently reaching for objects: “If I have my hands tied… it’s really helpful if I can just verbally let it know which [shelf] I want out.” A system with flexible interaction modes may be able to better accommodate these variable priorities.

\pagebreak
\begin{figure}[t]
  \centering
  \includegraphics[width=8.45cm]{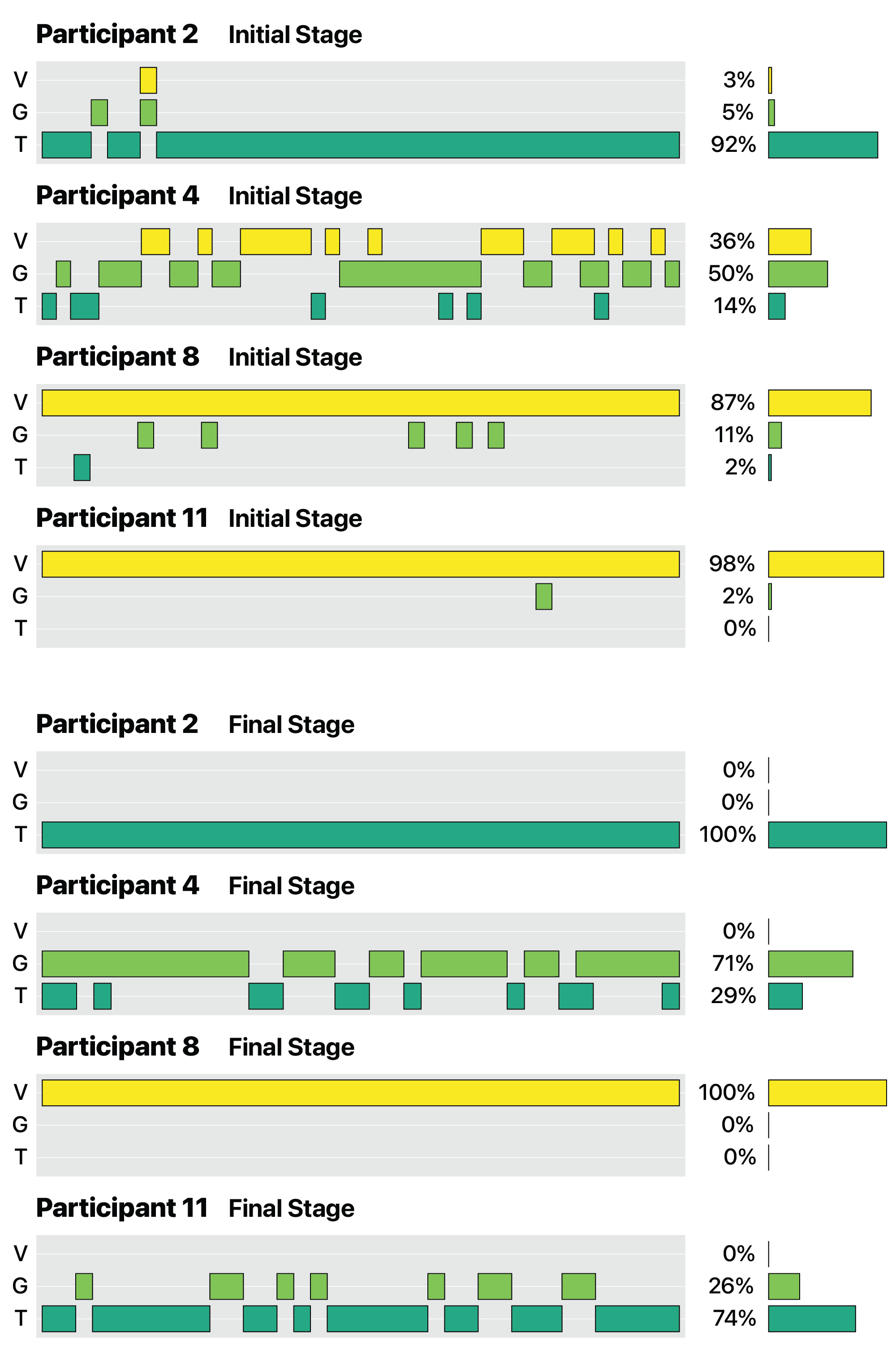}
  \caption{Timeline of vocalizations (V), gestures (G), and touch actions (T) for four participants (P2, P4, P8, P11), during both the initial and final stages of the study. These participants ended the study favoring either vocal actions or physical actions exclusively. The right side of the figure displays the distribution of actions for each participant and stage.}
  \Description{Timeline of vocalizations (V), gestures (G), and touch actions (T) for four participants (P2, P4, P8, P11), during both the initial and final stages of the study. These participants ended the study favoring either vocal actions or physical actions exclusively. The right side of the figure displays the distribution of actions for each participant and stage.}
\end{figure}

\noindent \textbf{Group 3: Single-Modal Preference.} The final group of participants (P2, P4, P8, P11) ended the study favoring one interaction mode — either purely vocal or purely physical (Figure 12). Within this group, only Participant 4 began the study using multiple modes. In the final stage, this participant eliminated vocalizations, explaining that these commands presented a larger cognitive load: “I have to say how many steps [the shelf] has to go down, and it involves some counting.” Participant 11, likewise, switched their primary mode from vocal to physical. The remaining participants maintained a consistent modal preference, rooted in technical assumptions about the capabilities of the system. “From the start, I already didn’t want to touch it,” said Participant 8, referencing concerns about mechanical stability. Participant 2 assumed the opposite: “I didn’t think about motion [sensing] as a possible factor,” they said, explaining why they opted for purely touch-based interactions instead of gestures. Notably, our forced-choice stages did not distinguish between gesture and touch (they were grouped into one “physical” category), which may explain this reasoning.

Overall, the majority of participants (8 out of 12) ended the study with a clear multimodal preference — demonstrated through recorded actions as well as debriefing interviews. It is worth noting that this multimodality was observed in two forms. The first is in the form of complementary actions — for instance, saying “make a shelf here” while touching a spot on the wall. The second is in the form of sequential actions — for instance, pointing at an out-of-reach object (gesture), then saying “move down” (voice), then retrieving the object, and then double-tapping the shelf in order to dismiss it (touch). As we will see in the following subsections, instead of deriving an explicit mapping from user actions to robotic responses, it can be helpful to interpret these actions as “evidence” towards a particular user intention, upon which a robotic system can act accordingly.

\subsection{Voice}

We recorded and transcribed a total of 848 vocalizations, over the course of the study. Some of these were simple, atomic commands (“higher”, “lower”), while others were more complex (“Column Three, I need a block placement at the same height as Column One”).

To better visualize and understand this dataset, we can embed the transcribed vocalizations in a two-dimensional feature space (Figures 12 and 13). Specifically, we use a pre-trained encoder model, based on MPNet \cite{song2020mpnet}, to compute a vector that captures the semantic meaning for each vocalization. Then, we apply t-SNE \cite{maaten2008visualizing} to reduce the dimensionality (from 768-D to 2-D), using a Euclidean distance metric.

Plotting these vocalizations, we can begin to examine the broad structure of the dataset. Vocalizations on the left side of the plot tend to be shorter — one to four words, usually describing a location (“here”, “there”), a short command (“go back”, “go up a bit”), or an anthropomorphic response \cite{kim2023anthropomorphic} (“there you go”, “good job”, “that’s fine”). Vocalizations are the right side of the plot tend be longer — they include requests (“can you go back to the straight status?”), detailed imperatives (“give me a curve on the second marker for the right column”), grounding statements related to the task (“I am a very short person, so…”), and multi-step instructions.

Notably, the clusters on the left of the plot, which represent shorter utterances, tend to contain vocalizations from a wider range of participants. This suggests a high degree of consensus for simple, atomic commands. Towards the right, as vocalizations become more complex, the clusters show less participant diversity — reflecting more individualized speech patterns. However, even within these areas, there is still a degree of semantic overlap, implying some shared strategies for articulating goals. We examine some of these patterns in the subsections below.

\begin{figure}[t]
    \centering
    \includegraphics[width=\linewidth]{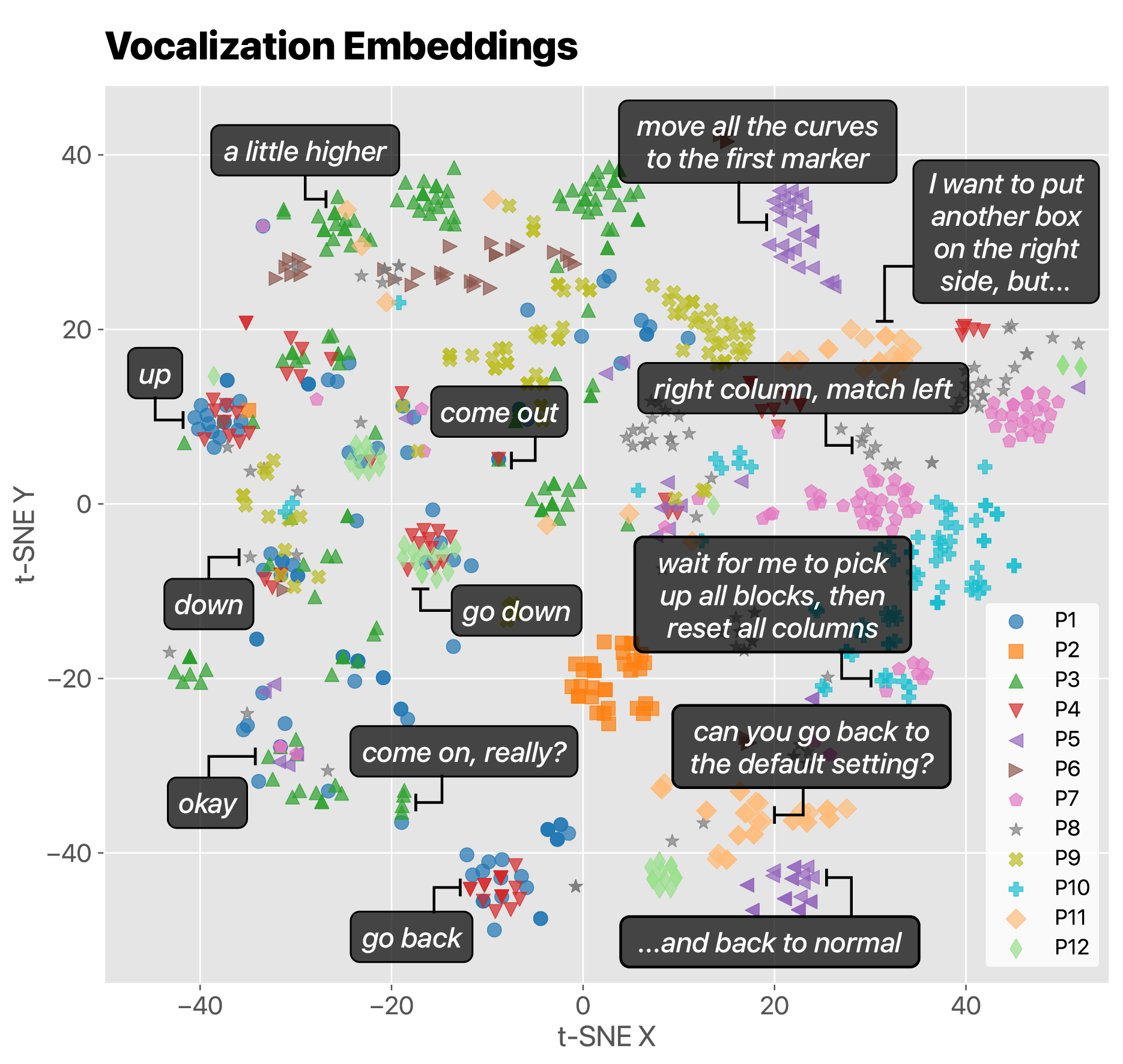}
    \caption{Semantic embeddings for participants' vocalizations during the elicitation study, visualized in a two-dimensional latent space. Vocalizations on the left of the plot tend to be shorter, and have higher overall agreement between participants. Vocalizations on the right of the plot tend to be longer, and reflect more individualized speaking styles.}
    \Description{Scatterplot showing participants’ vocalizations mapped in a two-dimensional latent space. Vocalizations on the left of the plot tend to be shorter and have higher overall agreement between participants. Vocalizations on the right of the plot tend to be longer and reflect more individualized speaking styles.}
\end{figure}

\begin{figure}[t]
\centering
\includegraphics[width=\linewidth]{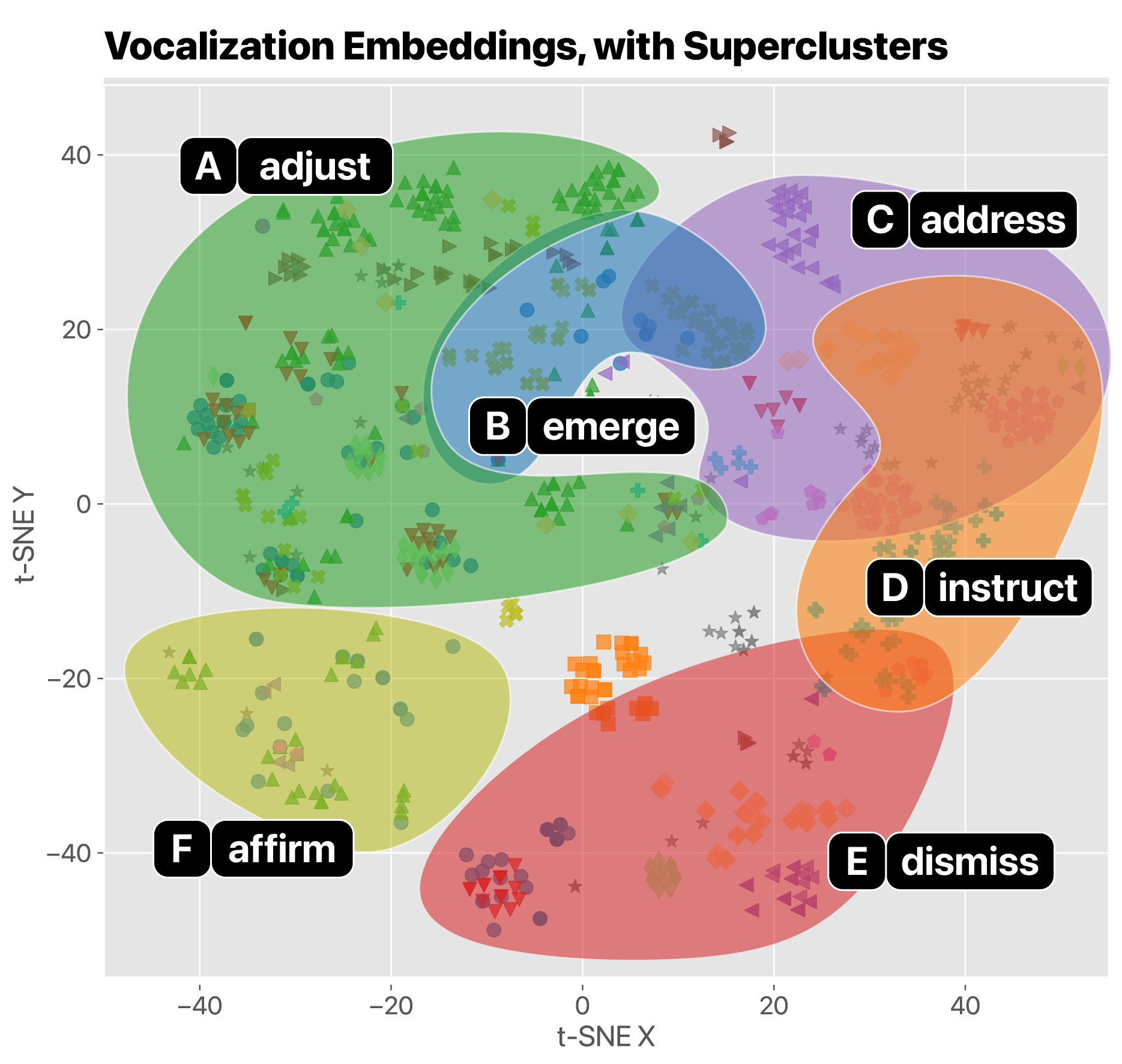}
\caption{Once participants' vocalizations are plotted, we can observe several superclusters, some of which overlap. Categories include simple instructions for vertical movement (a), commands for shelves to emerge (b) and retract (e), short anthropomorphic reactions to establish common ground (f), and longer, multi-step instructions (d), many of which were addressed (c) to specific columns on the wall.}
\Description{The same plot as Figure 13, but with five superclusters highlighted. These clusters correspond to different types of vocalizations, including vertical movement commands, shelves emerging or retracting, anthropomorphic reactions, multi-step instructions, and phrases addressed to specific columns.}
\end{figure}

\subsubsection{Manipulation Analogs.}
Many of the shorter voice commands can be interpreted as coarse substitutions for direct manipulation techniques. Most often, these were used for adjusting the height of a shelf: participants would give instructions such as “a little higher up” or “move all the way down”. Some participants expressed frustration at the imprecision of vocal manipulation, especially compared to more direct, physical methods. Others, however, were generally unbothered by the lack of fine control. These participants resolved uncertainties by referencing spatial landmarks (“go to the second notch from the bottom”) or building off of previous forms (“right column, match left”). To adequately respond to these commands, a system will have to incorporate knowledge of its physical surroundings as well as its physical state.

\subsubsection{Addressing and Implicit Context.}
Our robotic wall consisted of three columns, and participants could create shelves from any one of the three. To clarify their intentions, many participants chose to address these columns by name (i.e. “left column, make the same shelf as the middle column, please”). However, this was not intuitive for all. In the voice-only stage, 38\% of “new shelf” commands were unaddressed, requiring our human facilitator to interpret the implicit context around them (i.e. gaze, body posture) in order to respond. Context was even more critical for “adjustment” commands, 70\% of which were unaddressed. In this latter case, the context was largely temporal — a more ambiguous “adjustment” command was typically preceded by a less ambiguous “new shelf” command.

We should also note that while this addressing scheme is manageable for users in a laboratory setting, it would likely become cumbersome when applied to wider surfaces. A vocally-controlled room-scale system will presumably need to leverage implicit human cues, both spatial and temporal, just as our facilitator did.

\subsubsection{Emergence, Dismissal, and Goals.}
There were two main approaches to creating new shelves. The first approach is analogous to “sculpting” — participants would request a shelf without a specified location, and then use the manipulation analogs described earlier to refine the form. The second approach was a “one-shot” technique — participants would specify both the shelf and its target location in a single utterance (“middle column, shelf at the topmost position”).

There was a similar dichotomy among dismissal commands. Most participants alternated between action-oriented directives (“retract”, “reset”) and state-based descriptions (“back to normal”, “default position”). However, some ventured further, chaining multiple operations into procedural recipes: “Bring all columns to the bottom, let me pick up the blocks, and then reset them all, once you detect that there's nothing on your wall”.

A few participants chose to articulate abstract goals rather than explicit commands (e.g. “I want to put another box on the rightmost side”). These participants would also reference prior states, saying things like “do the same thing as before” or “return to the previous position.” Such commands assume the system maintains a working memory of past configurations.

\begin{table*}[h]
  \caption{Example Codes and Corresponding Descriptions (Gesture and Touch)}
  \begin{tabular}{ccl}
    \toprule
    Code&Description\\
    \midrule
    \verb|g_point_location| & "the user pointed to a panel on the wall" \\
    \verb|g_meet_object| & "the user held an object above or below a shelf on the wall" \\
    \verb|t_push_exterior_low| & "the user pushed the bottom of the shelf upward" \\
    \verb|t_tap_face_repeated| & "the user tapped a panel on the flat part of the wall, repeatedly" \\
    \verb|...| & ... \\
  \bottomrule
\end{tabular}
\end{table*}

\subsubsection{Establishing Common Ground.}
Throughout the study, participants actively worked to establish mutual understanding with the robotic wall. Confirmatory utterances like “okay, good” and “there you go” served as real-time feedback, signaling when the wall had correctly interpreted their intent. Conversely, when errors occurred, participants would verbally chastise the wall (“no, wrong”) or express frustration (“come on, really?”). This behavior was moderated by social context: one participant admitted they would be “less polite” to the wall in private settings, while others felt self-conscious speaking to an inanimate surface in a public setting.

Several participants attempted to construct a shared vocabulary, assigning names to specific heights (e.g. “alpha”, “bravo”) or numbering shelf locations for future reference. One participant even tried to establish persistent behavioral rules (“If there is nothing on the wall, reset the column”), seeking to offload cognitive burden onto the system itself. During debriefing interviews, participants emphasized the importance of common ground — questioning whether “left” referred to their perspective or the wall’s, and expressing a desire for explicit confirmation that the system was actively listening.

\begin{figure}[h]
  \centering
  \includegraphics[width=8.45cm]{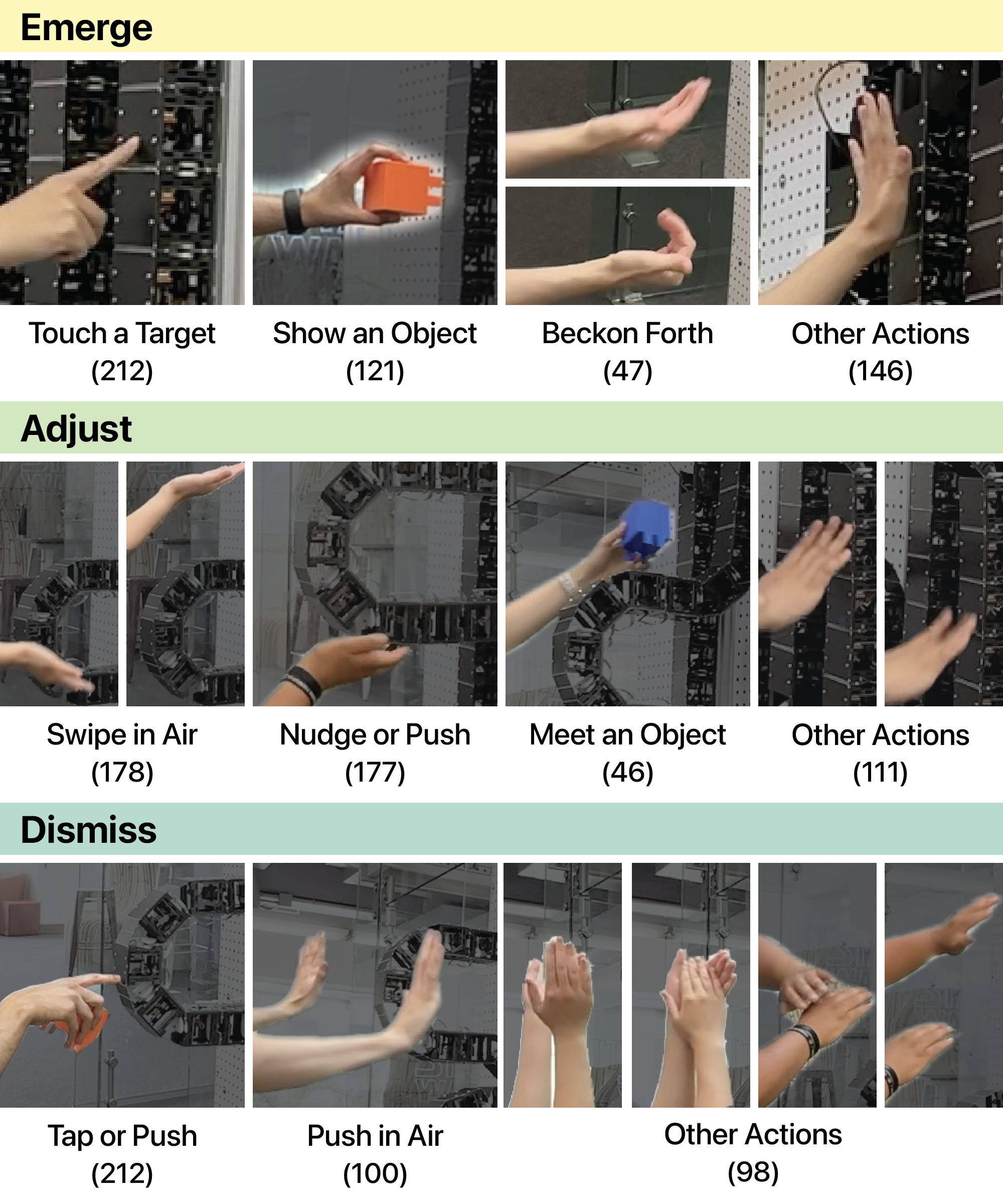}
  \caption{Pictured above are the most common gesture and touch actions that appeared in our elicitation studies. Raw counts are in parentheses, below each image.}
  \Description{Photographs of gesture and touch actions performed during the elicitation study.}
\end{figure}

\subsection{Gesture and Touch}

In our dataset, we identified 755 instances of gestures, and 724 touch actions. We define touch actions as hand or arm movements that involve physical contact with the robot wall (tapping, grabbing, pushing, etc.). All other communicative upper body movements were categorized as gestures.

The most common gesture and touch actions are summarized in Figure 15. Broadly speaking, touch actions were more often used for requesting and dismissing shelves, whereas for height adjustments, there was a more even split between modes. Beyond these functional differences, gestures were also more prevalent in multimodal interactions. In the first and final stages (i.e. free choice), 210 vocalizations were accompanied by gestures, whereas only 41 vocalizations were accompanied by touch actions.

We can expand this analysis by plotting these actions in a lower-dimensional feature space, similar to the one described in Section 4.4. Recall that each unique gesture and touch action is tagged with an appropriate subcode. These subcodes, however, can be expanded into descriptive sentences (as shown in Table 1), which we can use to perform a semantic analysis (Figures 15 and 16). This allows us to, at a glance, observe the agreement between participants for individual actions, as well as reason about larger behavioral patterns.  It is also useful for calculating a modal entropy, described in Section 4.6.

As we move up the vertical axis of Figure 16, actions become more “tactile”. At the bottom, we find mid-air sweeping motions and representational gestures performed at a distance from the wall. The middle zone contains gestures that approach but don't quite contact the surface — hands hovering near panels, objects held against shelves, and mimetic pushing motions. The top third consists almost entirely of direct touch actions, with the left cluster focusing on shelf manipulation and the right on interactions with the flat wall surface. We examine some of these clusters below.

\subsubsection{Object Recognition and Response.}
Several participants expected the wall to recognize and respond to objects held in space. They would bring objects towards the flat surface, or suspend blocks at desired shelf heights, waiting for the wall to “meet” the object with an appropriately positioned shelf. Participant 12 mentioned this explicitly: “The wall should basically follow me,” they said, “start slowly moving, so I know it’s following.”

\begin{figure}[t]
\centering
        \includegraphics[width=\linewidth]{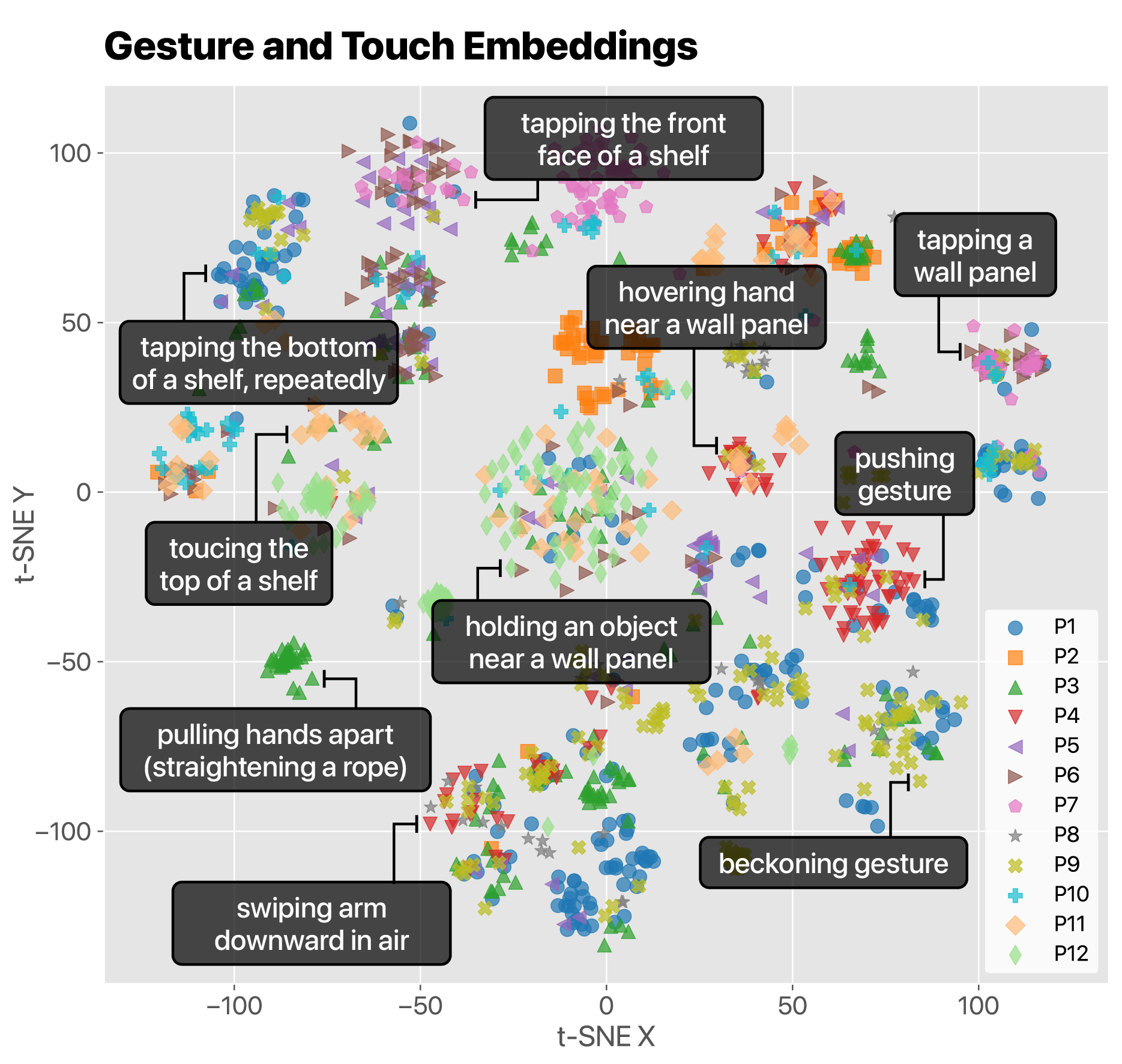}
        \caption{By attaching a text description to each gesture and touch action, we can compute semantic embeddings and visualize them in a two-dimensional latent space. Actions at the top of the plot are more tactile, involving direct manipulation with the wall. Actions near the bottom are gestures performed without contacting the robotic surface. In between are gestures that approach the surface, but don't quite touch it (a pushing gesture beside a deployed shelf, or a hand hovering over a flat panel).}
        \Description{A scatter plot of gesture and touch actions embedded in a two-dimensional latent space.}
\end{figure}

\begin{figure}[t]
\centering
        \centering
        \includegraphics[width=\linewidth]{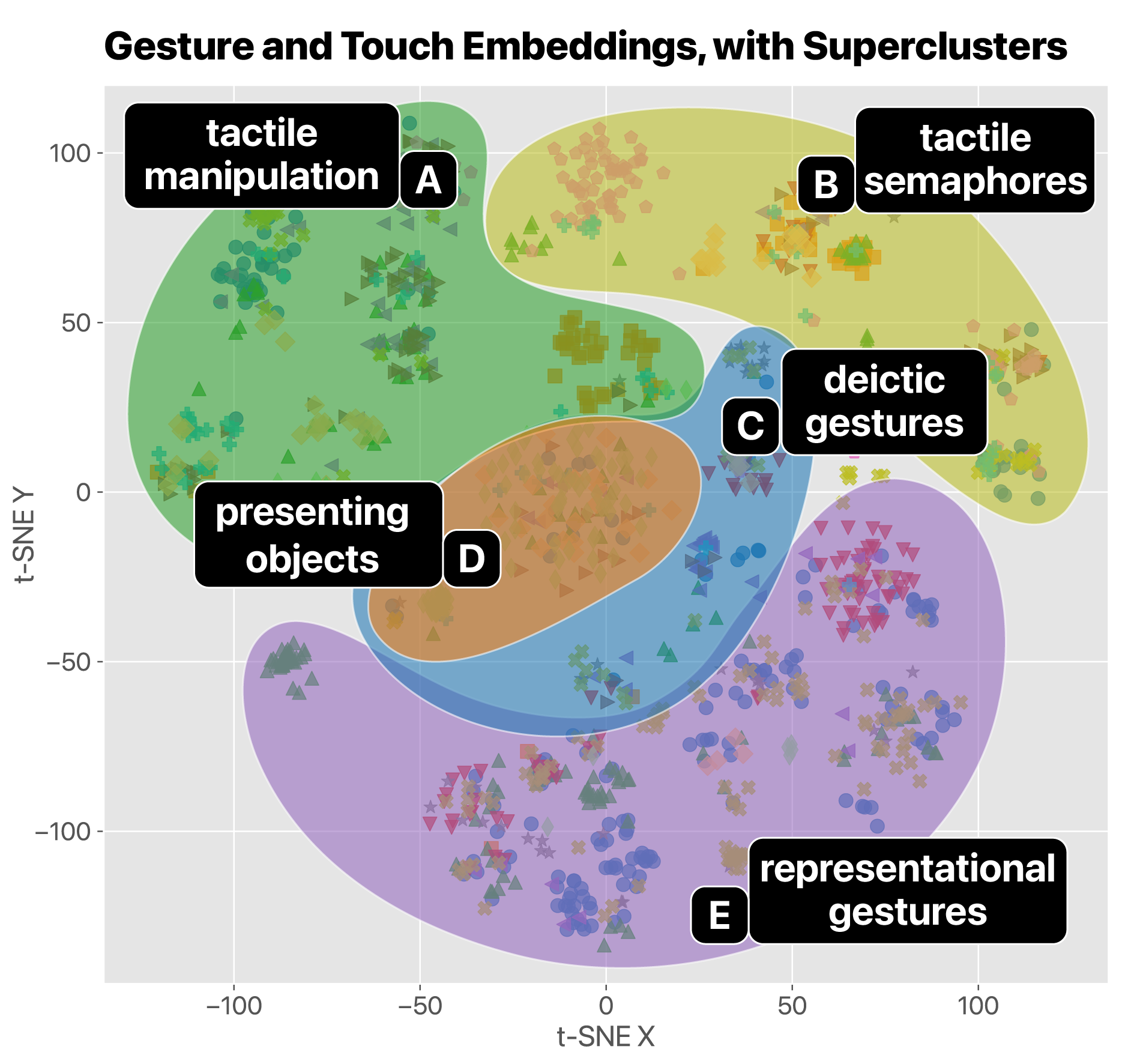}
        \caption{When gesture and touch actions are plotted, several superclusters appear. Categories include tactile "sculpting" actions (a), semaphoric tactile actions like taps and knocks (b), a variety of iconic and metaphoric gestures (e), and deictic gestures such as pointing (c). A subset of deictic gestures are interactions in which users which present objects to the wall (d) — holding blocks near certain locations, where they desired shelves to appear.}
        \Description{The same plot as Figure 16, but with superclusters highlighted to show categories of gestures, including tactile manipulation, tactile semaphores, deictic gestures, presenting objects, and representational gestures.}
\end{figure}

\subsubsection{Direct Manipulation.}
Touch interactions clustered into two behavioral patterns. The first involved direct physical manipulation of shelves and columns — participants often pushed, nudged, and tapped shelf edges to adjust heights and dismiss unwanted forms. Some attempted to “pull” shelves out of the wall in the flat state (though due to mechanical constraints, our facilitator had to discourage this). These actions leveraged the physical affordances of the robotic wall. A small number of participants instead borrowed actions from touchscreen vocabularies —  “scrolling” flat areas of the wall to alter shelf heights, or pinching the flat surface to solicit new forms.

The most common method of soliciting new shelves was to tap or touch a location — in the physical-only stage, 7 out of 12 participants used this as their primary technique. The remaining 5 participants used a combination of deictic and representational gestures, which we describe below.

\subsubsection{Representational Gestures.}
Some participants used iconic and metaphoric gestures \cite{studdert1994hand} to communicate with the wall. The most frequent choice was a “beckoning” gesture — to make shelves emerge, these participants would curl their fingers inward, or sweep their hands towards their body. One participant also attempted to command the wall through hand signs – holding up a number of fingers (one, two, or three) to signify that they were addressing a particular column.

A small number of participants also developed gestures for confirmation and control. Many of these followed a swiping or pointing gesture — for instance, a participant would motion for a shelf to move upward or downward, then hold a fist or put their palms outward when they wanted the wall to stop. When the wall behaved appropriately, some would give a “thumbs-up” sign. One participant also ventured into mimetic actions – pulling their arms apart (as if straightening a rope) to tell certain columns to “straighten  out” and retract.

\subsubsection{Deictic Gestures.}
In the free-choice stages, deictic gestures (specifically pointing) were often used in conjunction with voice commands. For soliciting shelves, 19\% of vocalizations were accompanied by a gesture, and for adjusting shelves, 44\% of vocalizations were accompanied by a gesture. Participants would gesture towards spots on the wall where they wanted a shelf to move or appear, or point up or down to indicate a direction.

While sitting, a few participants performed grabbing or reaching gestures to indicate that they wanted to retrieve an object that was located on a high shelf. Instead of directly telling the wall to move down, they expected the wall to interpret this action.

\subsection{Modal Entropy}

The two-dimensional embeddings in Sections 4.4 and 4.5 provide a helpful way to visualize and reason about the diverse interaction styles that appeared in our study. To further quantify this, we can interpret these plots as latent state spaces, and estimate the distribution of a user’s actions across both vocal and physical modalities. We introduce two metrics: an \emph{across-mode entropy} that captures the balance of user activity between different modalities, and a \emph{within-mode entropy} that describes the variety of user behaviors for any single interaction mode.

To begin, we compute two probability density functions for each participant. We achieve this through Kernel Density Estimation — within each latent space, we apply a Gaussian kernel to the participant’s two-dimensional action embeddings (Figure 18). The kernel bandwidth is unique for each participant, and is calculated via Scott’s Rule \cite{scott2015multivariate}, scaled by a factor of 0.4 to better capture the local structure of the clusters.

\begin{figure}[t]
  \centering
  \includegraphics[width=8.45cm]{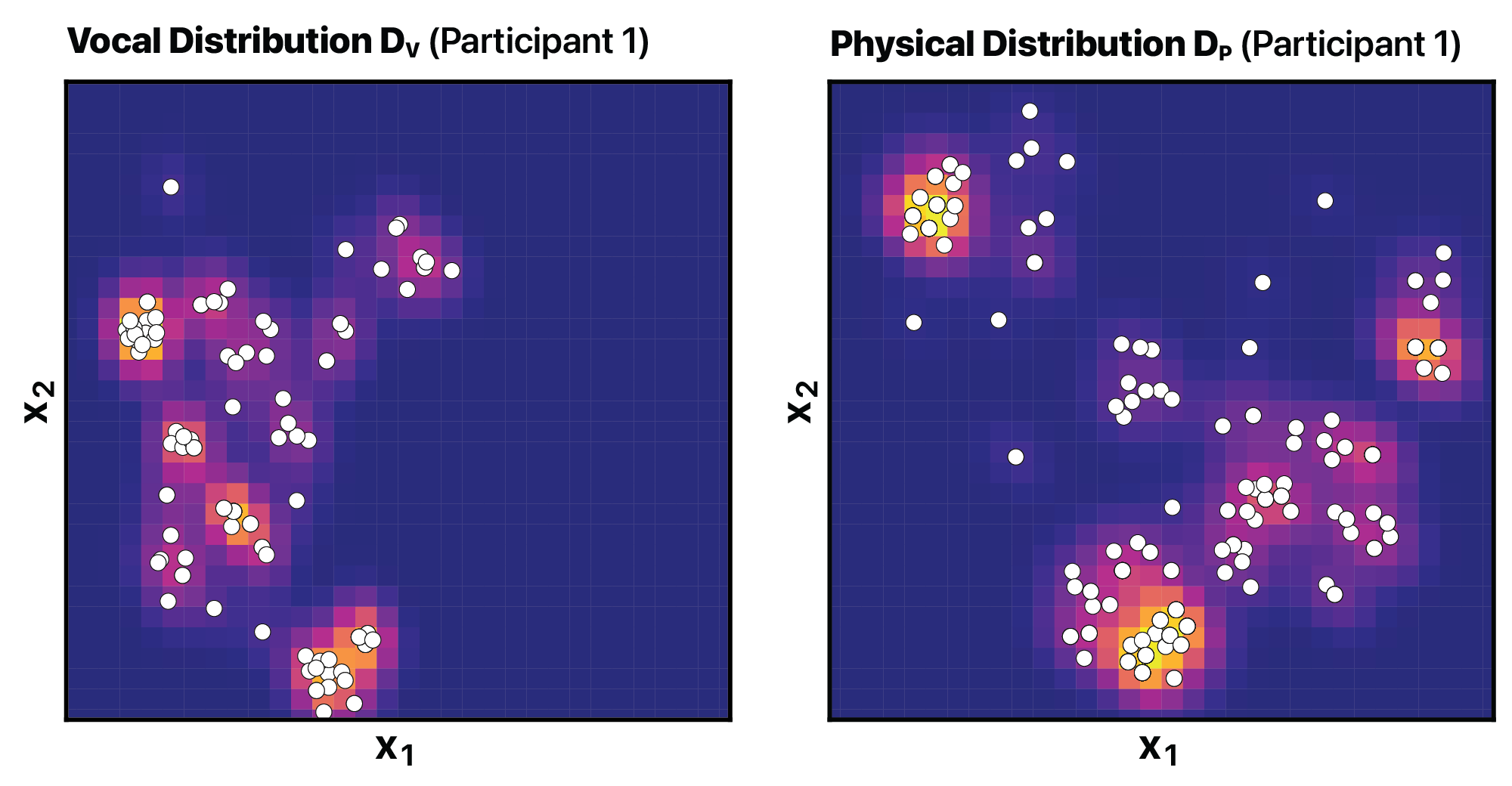}
  \caption{We can discretize the latent spaces in Figures X and Y, and interpret them as two non-overlapping state spaces. After applying a Gaussian kernel to participant actions within these spaces, we obtain probability mass functions for each participant’s vocal (a) and physical (b) actions. The plots above show distributions for one participant (P1), across all stages. White circles represent actual, observed actions.}
  \Description{Heatmap-like plots showing two state spaces, one for vocal actions and one for physical actions.}
\end{figure}

This process gives us one distribution for a participant’s vocal actions, $D_V$, and one for their physical actions, $D_P$. Formally, we can express this as:

\begin{equation}
    D(\textbf{x}) = \frac{1}{n}\sum_{i=1}^{n}K_{h}(\textbf{x}-\textbf{x}_{i})
\end{equation}

In the expression above, $\textbf{x}$ refers to a coordinate in the two-dimensional latent space, $n$ is the total number of actions for a participant within that space, and $K_{h}$ is the Gaussian kernel, with bandwidth $h$.

After obtaining the distributions $D_V$ and $D_P$, we can discretize them, evaluating each function on a uniform 32-by-32 grid. For each grid cell $k$, we compute $D(\textbf{x}_k)$ (where $\textbf{x}_k$ is the coordinate at the center of cell $k$) and multiply this value by the cell area:

\begin{equation}
    p_k=D(\textbf{x}_k) \cdot \Delta A
\end{equation}

After normalizing across the 1024 cells, we can then calculate the Shannon entropy for each modality:

\begin{equation}
    H=-\sum_{k=1}^{m}p_k\log_2(p_k) \ , \ m=1024
\end{equation}

This gives us an entropy $H_V$ for vocal actions, and an entropy $H_P$ for physical actions. Such metrics can help us distinguish between interaction styles that, at a coarse view, may appear to be similar. For instance, in the final stage of the study, Participants 1 and 12 both used physical actions around 80\% of the time (78\% for P1, 81\% for P12). However, the physical action entropy for P1 is nearly double that of P12 — reflecting their wide use of representational gestures, tactile manipulation, deictics, and so on.

We can further characterize a participant's interaction style by estimating their overall probability of performing a vocal or physical action. For a given number of vocal actions $n_V$ and physical actions $n_P$ in a free-choice stage, we obtain:

\begin{equation}
    \hat{P}_{V} = \frac{n_V}{n_V+ n_P} \ , \ \hat{P}_{P} = \frac{n_P}{n_V+ n_P}
\end{equation}

Then, we can use these estimations to calculate an \emph{across-mode entropy}, capturing the uncertainty surrounding a user's modality choice:

\begin{equation}
    H_{across}=-\hat{P}_{V}\ \text{log}_{2}(\hat{P}_{V}) - \hat{P}_{P}\ \text{log}_{2}(\hat{P}_{P})
\end{equation}

If a participant only uses vocal actions, or only uses physical actions, then their across-mode entropy will be zero. If they frequently alter between modes, then this entropy will increase.

Figure 19a plots the across-mode entropy for participants in both the initial and final stages of the study. Three clusters appear, which map directly to the qualitative groupings described in Section 4.3. Participants in the top left quadrant, for instance, began the study with a low across-mode entropy (using only physical actions), and ended the study with a high across-mode entropy (mixing speech, gesture, and touch). 

It can also be informative to derive a singular \emph{within-mode entropy} that captures both the vocal ($H_V$) and physical ($H_P$) entropies computed in Equation 3. For this, we use a simple weighted sum:

\begin{equation}
    H_{within} = \hat{P}_{V} \cdot H_{V} + \hat{P}_{P} \cdot H_{P}
\end{equation}

Figure 19b compares the across-mode and within-mode entropies for participants in the final stage of the study, yielding additional insights. Participants in the top-left quadrant stuck to one modality, but had a rich repertoire of actions within that modality (P8, for instance, was a "vocal specialist"). Participants in the bottom-right quadrant were the opposite — while they frequently switched between vocal and physical modes, their library of actions within each mode was more limited.

These metrics are most meaningful when used to compare participants within a population. To facilitate these comparisons, the entropy values plotted in Figure 19 are standardized (converted to z-scores). For a participant $j$, we can calculate the standardized across-mode entropy $z(H_{a}^j)$ and within-mode entropy $z(H_{w}^j)$, using the mean and standard deviation of the participant entropies in the population:

\begin{equation}
    z(H_{a}^j) = \frac{H_{a}^j-\mu_{H_{a}}}{\sigma_{H_{a}}} \ ,\ 
    z(H_{w}^j) = \frac{H_{w}^j-\mu_{H_{w}}}{\sigma_{H_{w}}}
\end{equation}

\begin{figure*}[t]
  \centering
  \includegraphics[width=12.0cm]{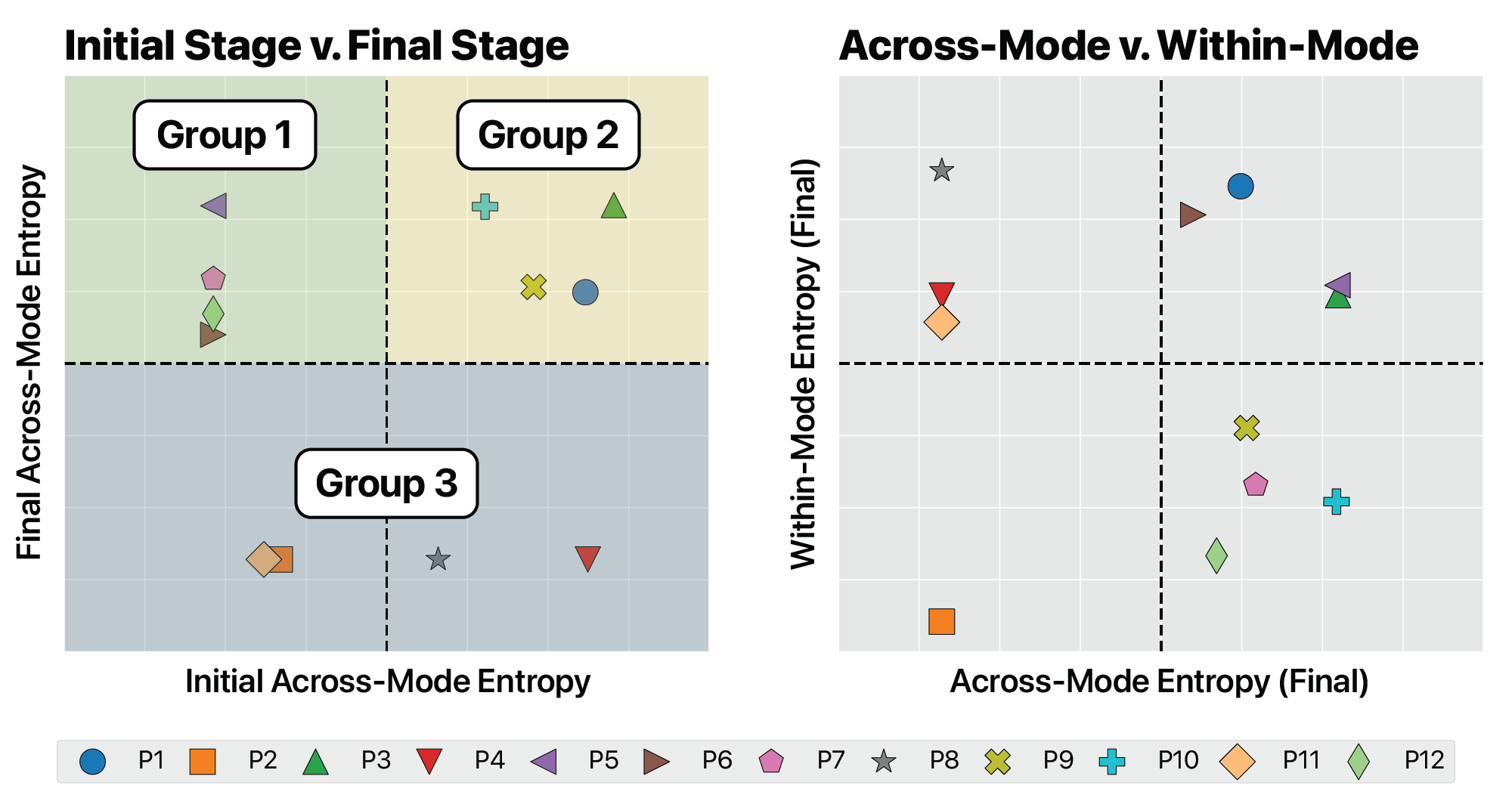}
  \caption{\textbf{Left:} Each participant's across-mode entropies for the initial and final stages are plotted against each other. Participants in Group 1 began the study using only one mode, but migrated to multimodal strategies by the end. Participants in Group 2 both began and ended the study using multimodal strategies. Participants in Group 3 began the study with varied styles, but all settled on a single interaction mode by the end. \textbf{Right:} Each participant's final across-mode entropy is plotted against their final within-mode entropy. A participant could have a lower across-mode entropy (left half of plot), but still have used a wide variety of actions within that mode (top half of plot).}
  \Description{Two scatterplots. Left: each participant’s across-mode entropies for the initial and final stages plotted against each other. Participants in Group 1 began using only one mode but migrated to multimodal strategies by the end. Participants in Group 2 both began and ended using multimodal strategies. Participants in Group 3 began with varied styles but all settled on a single interaction mode by the end. Right: each participant’s final across-mode entropy plotted against their final within-mode entropy. A participant could have a lower across-mode entropy but still have used a wide variety of actions within that mode.}
\end{figure*}

It is important to note that these measures are sensitive to the dimensionality reduction technique. In particular, the t-SNE embeddings that we use are dependent on  hyperparameters, which must be tuned in order to balance the preservation of local and global structures in the data. Despite this dependency, we found that a parameter sweep (of perplexity values ranging from 10 to 100) did not meaningfully alter the final groupings that appear in Figure 19. When comparing across-mode entropy to within-mode entropy, participants — with one exception — remained in the consistent quadrants for all parameter values.\footnote{For additional figures and source code, please see supplementary materials.} When comparing initial across-mode entropy to final across-mode entropy, participant positions are entirely unchanged, as this is independent of the t-SNE embeddings.

While we found that we were able to obtain descriptive results with this method, future researchers may also wish to consider TriMap \cite{amid2019trimap} or PaCMAP \cite{wang2021understanding}, which have been proposed as more robust to changes in parameters \cite{huang2022towards}. Overall, we see the metrics as useful diagnostic tools for coarsely identifying interaction styles within the context of a well-defined task (see Section 5.2).

\section{Discussion}

Cyber-physical architecture occupies a distinct niche within the broader robotic ecosystem. Unlike field or industrial robots, designers do not often imagine shape-changing walls as \emph{autonomous} — executing a set of goals, largely independent of the humans around them. However, they do envision them as \emph{agentic} — able to respond intelligently, or even proactively, to human needs and environmental triggers. Even in situations that are purely user-initiated, these robotic surfaces are expected to react in ways that leverage implicit context and overcome unstated instructions. Since these robots are one with the room, after all, designers perhaps assume that the robotic room “already knows” what’s happening within it.

This notion was reinforced by the results of our elicitation study. When debriefed (and sometimes during the study itself), users often mentioned that they would like the robotic wall to infer their intentions from the current context. “If I’m sitting,” said one participant, “it should always [create a shelf] at the lowest level.” Importantly, this inference goes beyond scene understanding — interpreting a user’s moment-to-moment interactions also requires knowledge of the temporal context, physical state, and co-occurrent actions (both vocal and physical). A simple mapping of human cues to robotic responses is likely insufficient for these tasks.

What, then, does this imply for the design of room-scale robotic interactions? When asked, users tend to \emph{say} that they desire a high level of control over these systems. However, in practice, they omit critical context and resist the cognitive burden of granular specification. This leads us to a bit of a paradox: In order to make these systems \emph{naturally controllable} for users, we may have to allow these systems to take actions that users \emph{do not explicitly command}.

\subsection{Understanding Implicit Context}

A key takeaway from our elicitation study is that natural interactions are seldom self-contained. Users issue commands that presuppose shared knowledge, rely on preceding exchanges, or assume the system can read situational cues. We observed this phenomenon in two forms: ambiguous signals and unstated desires.

\subsubsection{Ambiguous Signals.}
The same physical action can carry multiple meanings depending on context, or personal interaction styles. Consider a simple gesture: a user extends their arm toward the wall with palm facing outward. How should an intelligent system interpret this?

\begin{enumerate}
    \item A shelf is moving upward, and the user wants it to stop.
    \item The user just retrieved an object, and they want the empty shelf to retract back into the wall.
    \item The user wants to summon a shelf at a location near their palm.
    \item A shelf is too low, and the user wants it to rise up towards their arm.
\end{enumerate}

We observed all four intentions over the course of the study. Our wizard could disambiguate because they possessed task knowledge (for instance, understanding where the participant was trying to place a block, or whether they were attempting to solicit or dismiss a shelf). An automated system would need to maintain a similar awareness in order to resolve identical gestures into distinct meanings.

\subsubsection{Unstated Desires}

Beyond interpreting ambiguous signals, a more demanding expectation is that the environment should sometimes act without any explicit prompt at all. This theme surfaced repeatedly in our design workshops, where participants envisioned surfaces that could catch an elderly occupant mid-fall, or gently nudge a child away from a hazard.

These proactive behaviors, which might appear to leverage ``common sense'' or reflect social graces, can reduce cognitive load \cite{ju2008design} (something that many workshop participants valued as well). But it also raises the stakes for misinterpretation. For example, one elicitation participant suggested that our robotic shelves should automatically retract once an object is removed. This might be a reasonable heuristic in some contexts, but not all — if a user lifts a plant from a shelf to water it at the sink, the shelf should remain in place, not disappear. Just as in the previous subsection, the system must leverage context to infer an underlying goal.

\subsection{Design Tensions}

Within the design workshops, participants held competing desires. Many wanted environments that could anticipate their needs, handling minor tasks without prompting. Others insisted on explicit, user-initiated control, fearing misinterpretation or unwanted surveillance. (Some participants did both). Attitudes toward predictability were similarly split. Dependable behavior mattered for high-stakes scenarios, but some participants were also drawn to emergent, slightly unpredictable actions that suggest a theory of mind.

A separate tension emerged around ownership. Personalization was welcome in private settings, but shared spaces raised harder questions. Designers will have to consider how personalization can remain meaningful while mediating between multiple stakeholders. This includes how boundaries are negotiated, whose preferences are prioritized, and how fairness is maintained in shared environments.

\subsection{Towards Models of User Intent}

The findings above point toward a reframing of interaction design for robotic environments. Rather than treating user actions as discrete commands that trigger predetermined responses, we propose interpreting them as cumulative evidence feeding a dynamic model of user intent.

Our elicitation study offers some concrete grounding for this perspective. Seventy percent of shelf-adjustment commands were unaddressed — participants did not specify which column they meant, relying entirely on context established by earlier actions. Users also punctuated sequences with brief utterances (e.g. ``good job'' or ``no, wrong'') that functioned as real-time feedback, confirming or correcting the system's interpretation, and attempting to shape how subsequent actions should be parsed. These patterns indicate that meaning accrues across exchanges: each gesture, phrase, or touch refines the evidence available for understanding what comes next.

Intent modeling itself is not new to HCI. Early work by Horvitz \cite{horvitz1988decision, horvitz1998lumiere} demonstrated Bayesian networks for inferring user goals, and gesture recognition systems have long employed Hidden Markov Models for temporal disambiguation \cite{wilson2002parametric}. Probabilistic methods have also proven effective for handling input uncertainties in pen-based and touch interfaces \cite{schwarz2010framework}. However, many existing systems either process modalities through separate pipelines, or employ end-to-end architectures that map sensor data directly to actions without maintaining persistent belief states. These designs risk overlooking the temporal dependencies and cross-modal cues that our study revealed as central to natural interaction.

The modal entropy metrics introduced in Section 4.6 offer one lens for characterizing interaction styles, which could in turn inform how a system calibrates its inference. By quantifying how users distribute their actions across and within modalities, designers can distinguish between individuals who favor a narrow repertoire, and those who fluidly explore multiple channels. A high-entropy user might benefit from a parsing scheme that is more flexible (even at the expense of some accuracy); while a low-entropy user might prefer stable responses to a smaller set of familiar commands. We note, however, that these metrics are descriptive rather than prescriptive. For a given task, they can help characterize one user's behavior relative to others; but the metrics themselves do not indicate how a system should respond.

Cyber-physical architecture is uniquely positioned to leverage these models, in that it is persistent and omnipresent — unlike a traditional interface, it can begin reasoning about user intentions before a user approaches it or calls it to action. A shape-changing wall, for instance, might notice a person repeatedly glancing at a high shelf while carrying a bag, building confidence in the latent intent to “retrieve something” before any direct interaction occurs. A well-calibrated system could offer help when confidence is high, while deferring to explicit commands when intentions remain ambiguous. Striking this balance will be important, in order to create environments that are both responsive and respectful of human agency.

\section{So, What Should You Build?}

A key aim of our work was to uncover well-defined research directions for advancing the state-of-the-art of shape-changing environments. Drawing on evidence from our design workshops and elicitation study, we propose five opportunities for future investigators.

\subsection{"Raise" a Robotic Environment}
Shape-changing environments are a bit of an alien concept, and while they excited the designers and architects in our workshops, the participants also recognized that such structures might initially be jarring. In addition to an unfamiliar physical presence, a newly installed robotic wall might feel invasive if it “knows too much” about a user right out-of-the-box. To build trust gradually, participants suggested a more developmental approach — one likened it to “training a pet”. The system would begin with deliberately limited capabilities, but progressively learn skills and user preferences through repeated interactions.

Behavioral patterns from our elicitation study suggest that users might find this trajectory intuitive. Many participants expanded their interaction repertoires over time, attempting to establish shared vocabularies and teach persistent rules. A “raisable” environment would formalize this co-evolution. While slower to get up and running, users might feel greater ownership over the final robotic behavior, as it would be something that they taught the system themselves \cite{chi2023people}.

Researchers that pursue this direction will have to tackle challenges in both technical feasibility (i.e. persistent memory, one-shot learning), and interaction design. Neural models for continual learning, while growing increasingly advanced \cite{wang2024comprehensive}, have still been shown to be unreliable in HRI contexts, and trust drops significantly when robots are forgetful \cite{ayub2025continual}. Users preferred teaching methods are also unexplored within this context — in particular, the question of how users could force the system to “un-learn” misaligned behaviors.

\subsection{Directly Malleable Surfaces}

It is important for users to maintain authority over robotic environments. Workshop participants explicitly valued a form of tactile control, imagining scenarios where they could manually lift office dividers or physically reprimand misbehaving surfaces: “You could just jerk it back to where you want it to be and be like: ‘No!’”. Users embraced this instinct in our elicitation study as well: during the final stage, the majority of participants (9 out of 12) used touch actions at least a quarter of the time. They pushed shelf edges to adjust heights, tapped surfaces to dismiss forms, and attempted to physically mold the wall.

A promising research direction involves surfaces that can be manually sculpted \cite{Gonzalez2025SculptableMesh}, transitioning between malleable states for user control, and rigid states for structural stability. Such systems will need to distinguish between deliberate reshaping and incidental contact. They may also need to infer a user’s global-state intentions from their localized manipulations, in order to more fluently accommodate user behavior.

\subsection{Crafting Legible Intent}
When we debriefed participants after our elicitation study, many brought up the importance of establishing common ground. In particular, because there would often be a small delay while our robotic wall reconfigured its internal constraints,  participants suggested that subtle changes in lighting might help convey that the wall was still “paying attention” to vocal or gestural commands. Concerns over legibility extended to physical motion as well. Early in the study, one participant flinched at an abrupt shelf deployment, remarking that “it felt like I was being attacked.”

While motion legibility is well-studied for robotic manipulators \cite{dragan2013legibility} and mobile platforms \cite{lichtenthaler2012influence}, architectural-scale systems face a different set of challenges. They are non-anthropomorphic (and can not establish joint attention through gaze \cite{huang2010joint}), and certain forms of communication (e.g. artificial speech) may not always be preferred \cite{luria2017comparing}. Findings from our design workshops echo the known trust-building benefits of transparent behavior \cite{bhaskara2020agent}, but for cyber-physical architecture, the preferred means of transparency must still be uncovered.

\subsection{Co-Design for Accessibility}

While our design workshops surfaced compelling accessibility scenarios (adjustable counters for wheelchair users, guidance systems for physical therapy), these concepts emerged from participants without lived experience of disability, or formal training in the area. To engage with these ideas more seriously, researchers should consider centering disabled users as co-designers of shape-changing environments.

Aspects of our elicitation study hint at some of the complexity ahead. Seated participants often preferred voice and gesture over touch, suggesting that proximity and physical capability reshape one’s interaction preferences. Still, these preliminary observations are not substitutes for direct engagement with disability communities, who bring essential expertise about their own needs, and the social dynamics of assistive technology \cite{nanavati2023physically, threatt2014assistive}.

Future work could employ participatory design methods to explore how shape-changing environments might serve households with multiple, intersecting access needs — for instance, a family where one member uses a wheelchair, another has low vision, and a child has sensory processing differences. Such co-design efforts must navigate the friction that our workshops revealed between personalization and shared ownership, ensuring that adaptations don't stigmatize or inconvenience other users. Instead, this research direction should embrace “universal design” \cite{hamraie2017building} — an environment that is built to be flexible from the ground up is often preferable to one that is dotted with retrofitted accommodations.

\subsection{Large-Scale, Load-Bearing Structures}

Deployable and configurable seating emerged as one of the most common applications across our design workshops. Participants envisioned on-demand benches in transit hubs, adaptive chairs for growing children, and collapsible furniture in small apartments. Yet our prototype, like most shape-changing surfaces, cannot support human body weight. Those that do, often trade structural stability for geometric resolution \cite{kinch2014encounters} or configuration speed \cite{je2021elevate}. Creating structures that are simultaneously dynamic, load-bearing, and scalable remains an open engineering challenge.

\section{Limitations}

Both of our studies involved relatively homogeneous groups of participants, which likely influenced the values, scenarios, and interaction techniques that emerged. The majority of participants were younger adults (all under 40 years old), and we did not systematically attempt to include people with varying abilities, incomes, or occupations. This limits the diversity of perspectives represented in our findings, which ultimately restricts the generalizability of our results. In particular, since our speculative workshops featured only designers and architects, the eleven values we identified are not guaranteed to be consistent across all populations.

Our elicitation study examined a simple task within a laboratory setting. While this allowed us to conduct a controlled experiment, we note that participants may have chosen different interaction techniques outside of this particular context. For instance, some participants mentioned that when speaking to the wall, the tone of their commands (both volume and politeness) would likely change between public and private settings.

Additionally, although our ``wizard'' experimenter faithfully followed the protocol outlined in Section 4.1, a different experimenter might have had alternative interpretations of some ambiguous participant actions. (This is difficult to avoid entirely, as is the case in most Wizard-of-Oz studies \cite{riek2012wizard}). Even for non-anthropomorphic robots such as ours, the ``personality'' performed by the experimenter can influence how participants respond \cite{spadafora2016designing}. We attempted to mitigate this restricting our experimenter to a set of simple, predefined motions, accessible via a graphical control panel. However, this set of motions is itself a design choice, and variations in speed or rhythm have may have communicated different capabilities (as demonstrated in prior work \cite{sirkin2015mechanical, knight2017get}).

Finally, we note that the study focused on initial interactions and did not examine long-term adaptation or learning over time. Just as overall usage patterns for social robots have been shown to vary across multi-month periods \cite{ostrowski2022mixed}, it is also conceivable that certain interaction techniques might change as the novelty of a particular modality wears off.

\section{Conclusion}

This paper has investigated how people might live and interact with walls and surfaces that can physically transform on command. Our speculative workshops surfaced a set of eleven values that designers and architects find compelling — while also exposing deep frictions between automated support and personal agency, and between individual customization and public ownership. Our elicitation study revealed that many users intuitively blend speech, gestures, and direct contact when communicating with these structures, often assuming the system will interpret unspoken situational cues without explicit instruction.

A consequential tension emerges from both studies: users expect context-dependent responsiveness, yet the interactions they find most natural leave critical information unstated. Many of the scenarios that our workshop participants envisioned, for instance, require these systems to maintain a degree of awareness before any explicit action occurs. Balancing this expectation with simultaneous concerns over privacy and personal autonomy is difficult. But ultimately, as shown through our elicitation study, a robotic environment that does \emph{not} make these inferences could end up being \emph{less} controllable from a user's perspective — as it may not be able to respond adequately to the interaction techniques that users intuitively choose.

\begin{acks}
The work in this paper was supported in part by the National Science Foundation under Award No. IIS-2443190 and Award No. IIS-2420434. Additional support was provided by the Frank-Ratchye STUDIO for Creative Inquiry and the Extended Reality Technology Center.
\end{acks}

\bibliographystyle{ACM-Reference-Format}
\bibliography{references}


\end{document}